\documentclass[a4paper]{article}
\usepackage{graphicx} 
\usepackage{bm}
\usepackage{amssymb}
\usepackage{amsmath}
\usepackage{bm}
\usepackage{mathrsfs}
\usepackage{hyperref}
\usepackage{makeidx}
\usepackage{color}
\usepackage{amsmath}
\usepackage{amssymb}
\usepackage{fancybox}
\usepackage{bm}
\usepackage{amsthm}
\makeindex
\newtheorem{definition}{Definition}[section]

\newtheorem{remark}[definition]{Remark}

\newcommand{\QED}{\nobreak \ifvmode \relax \else
      \ifdim\lastskip<1.5em \hskip-\lastskip
      \hskip1.5em plus0em minus0.5em \fi \nobreak
      \vrule height0.75em width0.5em depth0.25em\fi}
\def\cP{{\hspace{-.1em}c\hspace{-.05em}\mathrm{P}}}

\def\SO{{\hspace{-.1em}S\hspace{-.1em}O\hspace{-.1em}}}
\def\SU{{\hspace{-.1em}Y\hspace{-.2em}M\hspace{-.1em}}}

\def\TM{{T\hspace{-.2em}\M}}
\def\TsM{{T^*\hspace{-.2em}\M}}
\def\TMM{{T\hspace{-.2em}\MM}}
\def\TsMM{{T^*\hspace{-.2em}\MM}}
\def\FF{{\mathcal F}}

\def\FFF{\mathfrak{F}}
\def\f{\mathscr{F}}
\def\wcs{{\www\hspace{-.1em}\cdot\hspace{-.1em}\sigma}}
\def\stars{{\star\star}}

\def\d.o.f\@.{{d.o.f$.$}}

\def\LLambda{\begin{large}\bm\Lambda\end{large}}

\def\bcdot{{\circ}}
\def\bcdots{{\circ\circ}}
\def\bcdott{{\circ\circ\circ\circ}}

\def\iii{{\mathfrak i}}
\def\cGR{{c_{\hspace{-.1em}g\hspace{-.1em}r}}}
\def\GSO{{G_{\hspace{-.1em}S\hspace{-.1em}O}}}

\def\ooo{\mathfrak{o}}

\def\Z{\mathbb{Z}}

\def\QQQ{\mathfrak{Q}}

\def\TTT{{\mathfrak T}}

\def\d{{\cal D}}

\def\td{d_{\www}}

\def\vvv{\mathfrak{v}}
\def\VVV{\mathfrak{V}}

\def\FFF{\mathfrak{F}}
\def\f{\mathscr{F}}

\def\ggg{\mathfrak{g}}

\def\sss{\mathfrak{s}}

\def\SSS{{\mathfrak S}}

\def\LLL{{\mathfrak L}}

\def\HH{{\cal H}}

\def\HHH{{\mathfrak H}}

\def\A{\mathscr{A}}
\def\aaa{{\mathfrak a}}
\def\AAA{{\mathfrak A}}

\def\BBB{{\mathfrak B}}
\def\bbb{{\mathfrak b}}

\def\CC{\mathbb{C}}

\def\CCC{{\mathfrak C}}
\def\ccc{{\mathfrak c}}

\def\Varepsilon{{\mathcal E}}

\def\eee{{\mathfrak e}}

\def\I{\mathscr{I}}

\def\M{{\cal M}}
\def\MM{\mathscr{M}}
\def\MMM{\mathfrak{M}}

\def\R{\mathbb{R}}
\def\Ri{R}
\def\RRR{\mathfrak{R}}

\def\www{{\mathfrak w}}

\def\delBRST{\delta_{\mathrm{\hspace{-.1em}B\hspace{-.1em}R\hspace{-.1em}S\hspace{-.1em}T}}}

\def\Tr{\mathrm{Tr}}

\def\SO{{\hspace{-.1em}S\hspace{-.1em}O\hspace{-.1em}}}
\def\LLLGF{{\LLL_{G\hspace{-.1em}F}}}
\def\LLLFP{{\LLL_{F\hspace{-.1em}P}}}
\def\LLLQG{{\LLL_{Q\hspace{-.1em}G}}}

\begin{document}
\title{Nakanishi--Kugo--Ojima quantization of general relativity \\in Heisenberg picture}
\author{Yoshimasa Kurihara\footnote{yoshimasa.kurihara@kek.jp}
\\
{\it\small High Energy Accelerator Organization (KEK), 
Tsukuba, Ibaraki 305-0801, Japan}
}
\date{}
\maketitle
\begin{abstract}
The Chern--Weil topological theory is applied to a classical formulation of general relativity in four-dimensional spacetime.
Einstein--Hilbert gravitational action is shown to be invariant with respect to a novel translation (co-translation) operator up to the total derivative; thus, a topological invariant of a second Chern class exists owing to Chern--Weil theory.
Using topological insight, fundamental forms can be introduced as a principal bundle of the spacetime manifold.
Canonical quantization of general relativity is performed in a Heisenberg picture using the Nakanishi--Kugo--Ojima formalism in which a complete set of quantum Lagrangian and BRST transformations including auxiliary and ghost fields is provided in a self-consistent manner.
An appropriate Hilbert space and physical states are introduced into the theory, and the positivity of these physical states and the unitarity of the transition matrix are ensured according to the Kugo--Ojima theorem.
The nonrenormalizability of quantum gravity is reconsidered under the formulation proposed herein.
\end{abstract}
\maketitle
\section{Introduction}\label{intro}
General relativity is among the fundamental theories concerning the spacetime structure of the Universe, and its correctness has been established by many experiments.
New evidence provided by the seminal discovery of gravitational waves in 2016\cite{PhysRevLett.116.241103}.
On the other hand, at the microscopic level, nature is described by quantum mechanics.
The standard theory of particle physics based on quantum field theory has been well-established through the discovery of the Higgs boson\cite{Aad:2012tfa,Chatrchyan201230}.
Thus, our understanding of nature covers a wide range of lengthscales, from the large-scale structure of the Universe to the microscopic behavior of subatomic particles.
However, these two fundamental theories (i.e., general relativity and quantum field theory) are not consistent.
The construction of a quantum theory of gravity is thus one of the fundamental goals of modern physics.

Herein, the ``quantum theory of gravity'' is understood as follows:
$1$) a theory that describes the behavior of four-dimensional spacetime in regions where the uncertainty principle is essential (i.e., the Planck length);
$2$) a theory consistent with well-established general relativity at large scales; and
$3$) a theory that can provide experimentally measurable predictions.
Following the establishment of general relativity and quantum mechanics during the 1920s, the development of quantum gravity began in the 1930s.
For a detailed history, see \cite{Rovelli:2000aw,doi:10.1142qg} and references therein.
Three main-streams of gravity quantization are considered\cite{Rovelli:2000aw}:
$1$) The covariant perturbative approach\cite{Fierz211}: 
following the successful application of the QED method, small fluctuations from flat Minkowski space are treated as perturbations, from which the Feynman rule of gravitational interaction is derived. 
This method became less widely used after the discovery of the nonrenormalizability of theories\cite{'tHooft:1974bx} but saw  a resurgence in use  following the development of the super-string theoretical approach;
$2$) canonical quantization of metric tensor\cite{DeWitt:1967yk,DeWitt:1967ub,DeWitt:1967uc}: 
the metric tensor is treated as a dynamic variable and interpreted as an operator, after which it is quantized using the canonical method that requires commutation relations. 
The quantum equation of motion is obtained as the Wheeler--de\hspace{.1em}Witte equation\cite{DeWitt:1967yk}. 
This approach is also less widely used at present because the Wheeler--de\hspace{.1em}Witte equation is not mathematically well-defined; recently, it has been renovated as loop-gravity\cite{rovelli2004}, and further developments are ongoing.
$3$) path-integration quantization:
when the path-integration method is applied simply to gravity, nonrenormalisable infinities appear similar to those in the first approach. 
Spin network\cite{Rovelli:1995ac} and spin foam\cite{Perez:2012wv} methods are categorized as such an approach.
A nonperturbative approach for path-integral quantization\cite{Barenz:2012av,Krishnan:2016tqj,hamber2008quantum} and its extension such as a causal dynamical triangulation methods\cite{Ambjorn:1995jt, Ambjorn:1995aw, Ambjorn:2013tki, PhysRevD.94.024058, Loll_2019, ambjorn2009geometry, carfora2012quantum} are actively investigated to date.
Despite decades of intensive efforts on quantization, a widely accepted theory satisfying the abovementioned requirements has not yet been developed.

Compared with the four-dimensional case, the Chern--Simons topological theory informs us that quantum gravity exists in $(1\hspace{-.2em}+\hspace{-.1em}2)$-dimensional space\cite{witten1988} that it is renormalisable, and that it does not exhibit dynamics\cite{Witten198846}.
Three-dimensional general relativity does not exhibit any dynamic degree of freedom, even at a classical level.
In \cite{Witten198846}, Witten mentioned two points: the existence of quantum gravity is because of an accidental feature of the three-dimensional case in which the invariant quadratic form exists, and the nonrenormalizability of quantum gravity in four-dimensions is essentially because of the absence of the invariant quadratic that allows us to understand short-distance limits of spacetime as trivial zero-energy solutions. 
Indeed, the Chern--Simons action can provide a topological invariant only in odd-dimensional space\cite{0264-9381-29-13-133001}. 

Although at first glance it seems challenging to construct quantum general relativity using the Chern--Simons form in four-dimensional spacetime, we discovered a novel symmetry, referred to as the co-Poincar\'{e} symmetry\cite{doi:10.1063/1.4990708}, which allows the construction of general relativity according to the Chern--Weil theory in four-dimensional spacetime.
This new symmetry is an extension of translation symmetry, which, when applied to a pure gravitational Lagrangian without the cosmological term, produces only a total derivative term.
(Alternative approaches to modify a translation operator in the Poincar\'{e} group are proposed, e.g., the metric-affine gauge theory of gravity\cite{Hehl:1976kj, Neeman:1978nwt, Hehl:1994ue}.)
In this study, we show that invariant quadratic can be defined in four-dimensional spacetime by introducing Lie algebra to the co-Poincar\'{e} group, such that the Einstein--Hilbert gravitational Lagrangian can be defined as a second Chern class, in which there is no cosmological term.
A missing piece of quantum gravity in the four-dimensional Chern--Simons thory, the invariant quadratic form, is discovered in the four-dimensional Chern--Weil theory.

Our approach to quantization is based on the second category of ``canonical quantization of the metric tensor'' in the abovementioned list of quantization approaches.
For this method, the subject to be quantized is not spacetime itself; thus, the spacetime coordinate $x^\mu$ is not the q-number (operator) but rather the c-number\cite{nakanishi1990covariant,NakanishiSK2009}.
The subject for quantization is the metric tensor $g^{(c)}_{\mu\nu}(x)$, which is obtained as a solution of the Einstein equation.
In classical general relativity, the geometrical metric tensor $g^{(g)}_{\mu\nu}(x)$ is given by the solution of the classical Einstein equation, such as $g^{(g)}_{\mu\nu}(x)=g^{(c)}_{\mu\nu}(x)$, i.e., it is Einstein's equivalence principle.
At the quantum level, this relation is not fulfilled simply, and the geometrical metric tensor is given as the expected value of the quantum metric tensor $g^{(g)}_{\mu\nu}(x)=\langle g^{(q)}_{\mu\nu}(x)\rangle$.
A space whose metric tensor is provided as the expected value of a stochastic processes is discussed by the author in \cite{Kurihara_2018}.

An outline of our method is as follows.
We start from the Einstein--Hilbert action of general relativity, which has co-Poincar\'{e} symmetry and a principal bundle induced by this symmetry.
Accordingly, the fundamental spin and surface forms can be identified; they will be defined in the following section.
Based on these fundamental forms, Nakanishi--Kugo--Ojima covariant quantization is performed, and the complete set of quantum Lagrangians, equations of motion, BRST transformations and charges\cite{Becchi:1974md,Tyutin:1975qk} for pure gravity is obtained.
As a consequence of quantization, the scattering matrix is shown to conform to the Kugo--Ojima theorem\cite{kugo1979local,Kugo1978459}.

In this article, section \ref{prep} introduces the mathematical preliminaries of differential geometry to explain our terminology and conventions.
A standard formalism of gravitational Lagrangian is introduced, after which the Hamiltonian formalism is discussed based on the symplectic structure of general relativity\cite{Kurihara_2020}.
In section \ref{cqg}, an explicit formulation of the canonical quantization of general relativity using Nakanishi--Kugo--Ojima formalism\cite{nakanishi1990covariant} is given. 
Section \ref{discussion} discusses how to construct an appropriate Hilbert space and physical states in it.
It is shown that the unitarity of the quantum gravitational $S$-matrix is ensured owing to the Kugo--Ojima theorem.
The renormalizability of quantum general relativity is also discussed in section \ref{discussion}.
A summary of this study is presented in section \ref{summary}.

\section{Preliminaries}\label{prep}
Standard classical general relativity is geometrically reformulated in terms of the vierbein formalism according to previous studies\cite{fre2012gravity,Kurihara2018,Kurihara_2020}.
Futhermore, Lagrangian and Hamiltonian formalisms can be introduced using the vierbein formalism.
\subsection{Differential geometry}\label{DG}
A four-dimensional pseudo-Riemannian manifold $(\MM,\bm{g})$ with $GL(4,\R)$ symmetry is considered, on which each coordinate patch $U_p\subset\MM$ around $p\in U_p$, coordinate vectors are introduced as $x^\mu$.
The signature of the metric $\bm{g}$ is assumed to negative.
Accordingly, we take standard base vectors in tangent space $T_p\MM$ as $\partial_\mu$, and those in cotangent space $T_p^*\MM$ as $dx^\mu$.
The abbreviation $\partial_\bullet:=\partial/\partial x^\bullet$ is used throughout this paper.
The entire manifold is covered by such coordinate patches, and the tangent space $\TMM=\bigcup_pT_p\MM$ and cotangent space $\TsMM=\bigcup_pT^*_p\MM$ are introduced as trivial bundles in $\MM$.
It is notable that these two base vectors are dual of each other such that $dx^\mu\partial_\nu=\delta^\mu_\nu$.
Einstein's convention for repeated indices is used throughout  this study. 
As noted above, for the purposes of this study, forms are represented using the Fraktur letters ``$\AAA,\aaa,\BBB,\bbb,\CCC,\ccc\cdots$''.

General relativity is constructed in Riemannian space $(\MM,\bm{g})$ and it is always possible to identify a frame in which the affine connection vanishes as $\Gamma^\rho_{~\mu\nu}(p)=0$ at any point $p\in\MM$, which has $SO(1,3)$ symmetry.
This local manifold at $p$ is referred to as the local Lorentz manifold and is denoted by $\M_p$.
A trivial bundle $\M:=\bigcup_p\M_p$ exists trivially as the principal bundle in $\MM$ with structural group $SO(1,3)$.
We adopt the metric tensor $diag[{\bm \eta}]=(1,-1,-1,-1)$.
The tangent and cotangent spaces $\TM$ and $\TsM$ are introduced together with those in $\MM$.
For the representation used herein, the base manifold of vectors and forms are distinguished by Greek suffixes in $\MM$ or Roman suffices in $\M$.
Map $\Varepsilon:\MM\rightarrow\M:{\bm g}\mapsto{\bm \eta}$ is represented using a standard coordinate as follows:
\begin{eqnarray}
\eta^{ab}&=&\Varepsilon^a_{\mu_1}(x)\Varepsilon^b_{\mu_2}(x)g^{\mu_1\mu_2}(x).
\end{eqnarray}
The function $\Varepsilon^a_{\mu}(x^\mu)=[{\bm \Varepsilon}(x^\mu)]^a_{\mu}$ is referred to as the vierbein.
Map $\Varepsilon$ induces homomorphism $\TsMM\cong \TsM$; thus, the form $\aaa\in\Omega^p(\TsM)$ and its pull-back $\Varepsilon^*[\aaa]\in\Omega^p(\TsMM)$ are equated to each other and denoted simply $\aaa\in\Omega^p$ herein.
Orthogonal base vectors in $\TM$ and $\TsM$ are provided from those in $\TMM$ and $\TsMM$ as $\partial_a:=\Varepsilon_a^\mu\partial_\mu$ and $\eee^a:=\Varepsilon^a_\mu dx^\mu$, respectively.
They are dual bases of each other such that
$
\eee^a\cdot\partial_b=\Varepsilon^a_{\mu_1}\Varepsilon_b^{\mu_2}dx^{\mu_1}\partial_{\mu_2}
=\Varepsilon^a_{\mu}\Varepsilon_b^{\mu}=\delta^a_b
$.
The base vector $\eee^a$ is referred to as the vierbein form.
The one-form object $\www$ is a connection-valued one-form in $\TsMM$ and is referred to as the spin form with respect to the structural group $SO(1,3)$, which satisfies a relation $\www^t\bm\eta+\bm\eta\www=0$.
The spin form can be represented using a standard basis as $\www^{ab}=\omega^{~a}_{\mu~c}\eta^{cb}dx^\mu$ where $\omega^{~a}_{\mu~c}:=[\www]^{~a}_{\mu~c}$.
This form is antisymmetric as $\www^{ab}=-\www^{ba}$, where $\omega^{~a}_{\mu~c}$ is referred to as the spin connection.
The covariant derivative for $p$-form $\aaa\in\Omega^p$ is defined as follows:
\begin{eqnarray}
\td\aaa&:=&d\aaa+\cGR\hspace{.1em}[\www,\aaa]_\wedge,\label{connec2}
\end{eqnarray}
where $[\www,\aaa]_\wedge=\www\wedge\aaa-(-1)^{p}\aaa\wedge\www$.
The real constant $\cGR\in\R$ is referred to as a gravitational coupling constant.
The Lorentz transformation ${\GSO}=Hom(\Omega^p,\Omega^p)$ of the vierbein and spin connection forms are respectively:
\begin{eqnarray}
{\GSO}:\eee\mapsto{\GSO}(\eee)=\LLambda\eee,&{\rm and}&
{\GSO}:\www\mapsto{\GSO}(\www)=
\LLambda\www\LLambda^{-1}+{\cGR}^{-1}\LLambda d\LLambda^{-1},\label{LT}
\end{eqnarray}
where $\LLambda$ is the matrix representation of ${\GSO}$.
Direct calculations show that the covariant derivative $\td$ defined above is invariant under the Lorentz transformation. 

The torsion two-form $\TTT^a$ is defined using the covariant derivative as follows:
\begin{eqnarray}
\TTT^a&:=&\left[\td\eee\right]^a=
\left[d\eee+\cGR\hspace{.1em}\www\wedge\eee\right]^a=
d\eee^a+\cGR\hspace{.1em}\www^a_{~b}\wedge\eee^b,
\end{eqnarray}
where the rightmost expression is a component expression using a standard basis.
Hereafter, dummy Roman-indices are often abbreviated as a circle (or star), whereas dummy Greek-indices are not abbreviated to clarify the base manifold of forms.
Where multiple circles appear in an expression, their pairing must occur in a left-to-right order for both upper and lower indices.
In this notation, the definition of the curvature two-form is represented as follows: 
\begin{eqnarray}
\RRR^{ab}&:=&\left[d\www+\cGR\hspace{.1em}\www\wedge\www\right]^{ab}=
d\www^{ab}+\cGR\hspace{.1em}\www^a_{~\bcdot}\wedge\www^{\bcdot b}. 
\end{eqnarray}
In standard definitions of covariant derivatives and their curvature form, the gravitational coupling constant $\cGR$ either does not appear or is simply set to unity.
When only gravitational interactions are considered, i.e., without any matter or gauge fields, this constant can be absorbed within the definition of spin connection and does not affect the discussions of classical general relativity.
However, when performing the simultaneous quantization of gravity and, for example, the Yang--Mills theory, this coupling constant provides deep insight into the quantized gravity.

It is notable that there are two tensors with constant values in $\TM$, i.e., the metric tensor  ${\bm \eta}$ and Levi Civita tensor ${\bm \epsilon}$; these are not constant in $\TMM$.
Their two-dimensional surface form is defined as 
\begin{eqnarray}
\SSS_{ab}:=\frac{1}{2}\epsilon_{ab\bcdots}\eee^\bcdot\wedge\eee^\bcdot,\label{surface}
\end{eqnarray}
which is a two-dimensional surface that is perpendicular to both $\eee^a$ and $\eee^b$, in which $\epsilon_{abcd}$ is a completely antisymmetric tensor (the Levi Civita tensor) whose component is $\epsilon_{0123}=1$ in $\TM$.
Whereas the spin-form $\www^{ab}$ does not conform to a $SO(1,3)$ tensor value as shown in equation (\ref{LT}), the curvature form $\RRR^{ab}$ is a Lorentz tensor-valued two-form.
More accurately, $\RRR^{ab}$ is a rank-$2$ tensor in $\TM$ with respect to Roman indices and is two-from in $\TsM$ (and also in $\TsMM$) such as $\RRR\in\Omega^2\otimes V^2(\TM)$.
The curvature form can be expanded using a standard basis in $\TsM$ as $\RRR^{ab}=:\Ri^{ab}_{~~\bcdots}~\eee^\bcdot\wedge\eee^\bcdot/2$, where $\Ri^{ab}_{~~cd}$ is referred to as the Riemann curvature tensor.
Another type of curvature is introduced {\it via} the contraction of indices for the Riemann curvature tensor such as $R_{ab}:=\Ri^{\star\bcdot}_{~~\star a}~\eta_{\hspace{.1em}\bcdot b}$ (the Ricci tensor) and $R:=\Ri^{\bcdots}_{~~\bcdots}$ (scalar curvature).
Two Bianchi-identities are provided as follows:
\begin{eqnarray}
\td\TTT^a=
\left[\cGR\hspace{.1em}\RRR\wedge\eee\right]^a=
\cGR\hspace{.1em}\RRR^{a\bcdot}\wedge\eee^{\bcdot}\hspace{.1em}\eta_{\bcdots},&{\rm and}&
\td\RRR=0.
\end{eqnarray}
$GL(4,\R)$ invariant volume form $\vvv$ is defined as
\begin{eqnarray}
\vvv:=\frac{1}{4!}\epsilon_{\bcdott}\eee^\bcdot\wedge\eee^\bcdot\wedge\eee^\bcdot\wedge\eee^\bcdot
=det[{\bm \Varepsilon}]dx^{0}\wedge dx^{1}\wedge dx^{2}\wedge dx^{3},\label{vvv}
\end{eqnarray}
where $\det[{\bm \Varepsilon}]=\sqrt{-det[{\bm g}]}$.

A new operator, termed a bar-dual operator, is introduced for shorter representations of long formulae as follows:
A map that transfers the rank-$p$ tensor to the rank-$(n\hspace{-.15em}-\hspace{-.15em}p)$ tensor is defined in an $n$-dimensional spacetime manifold according to
\begin{eqnarray}
\overline{\aaa^{a_1\cdots a_p}}&=&\overline{\aaa}_{a_1\cdots a_{n-p}}=~~
\frac{1}{p!}\epsilon_{a_1\cdots a_{n-p}b_1\cdots b_p}
\aaa^{b_1\cdots b_p},\\
\overline{\aaa_{a_1\cdots a_p}}&=&\overline{\aaa}^{a_1\cdots a_{n-p}}=-\frac{1}{p!}
\epsilon^{ b_1\cdots b_pa_1\cdots a_{n-p}}
\aaa_{b_1\cdots b_p},
\end{eqnarray}
where $\aaa\in\Omega^p$ and $\bm\epsilon$ are completely antisymmetric tensors in $n$-dimensional space.
Thus, $\overline{\overline{\aaa}}=\aaa$ when $\aaa$ is antisymmetric with respect to all indices. 
The surface form can be expressed using this expression as $\SSS_{ab}=\overline{\eee^a\wedge\eee^b}$.
In contrast to the Hodge-$*$ operator (which maps the $p$-form to the $(n\hspace{-.2em}-\hspace{-.2em}p)$-form), this entirely distinct operator maps the rank-$p$ tensor to the rank-$(n\hspace{-.2em}-\hspace{-.2em}p)$ tensor while keeping the rank of the form as $\bar{\bullet}:\Omega^q\otimes V^p(\TM)\rightarrow\Omega^q\otimes V^{n-p}(\TM)$.
The Hodge-$*$ operator is not appropriate for constructing general relativity because the metric tensor is hidden within its definition.

\subsection{Lagrangian formalism}\label{lagrange}
The Lagrangian form of gravity that reproduces the Einstein equation without matter and gauge fields is written as follows:
\begin{eqnarray}
\LLL_G(\www,\eee)&=&
\frac{1}{\hbar\kappa}\left(
\frac{1}{2}\RRR^\bcdots\wedge\SSS_\bcdots-\Lambda_c\vvv
\right),\label{Lagrangian44}
\end{eqnarray}
where $\Lambda_c$ is the cosmological constant and $\kappa=4\pi G/c^4$ is the Einstein gravitational constant.
This Lagrangian commonly known as the Einstein--Hilbert Lagrangian.
The gravitational action is introduced as the integration of the Lagrangian such that
\begin{eqnarray}
\I_G
&=&\int_{\Sigma}\LLL_G.\label{gravact}
\end{eqnarray}
For example, the integration region $\Sigma$ is taken to be the entire space of $\MM$.
The speed of light is set to unity (i.e., $c=1$), the gravitational constant $\kappa$ has a value of $\kappa\approx 1.03829 \mathrm{s}^2\mathrm{m}^{-1}\mathrm{kg}^{-1}$, and the reduced Plank-constant $\hbar=h/2\pi$ is written explicitly.
Using these units, dimensions of physical constants are given as $[\hbar\hspace{.1em}\kappa]=L^2=T^2$ and $[\hbar/\kappa]=E^2=M^2$, where $L$, $T$, $E$, and $M$ are the length, time, energy, and mass dimensions. 
The cosmological constant has the length inverse-squared, $[\Lambda_c]=L^{-2}$, in the units used herein.
The Lagrangian form and action integral are set to be null-dimensional object $[\LLL_G]=[\I_G]=1$.
Although, this Lagrangian form is invariant under the general coordinate and local ${SO}(1,3)$ transformations, it is not invariant under the local translation\cite{0264-9381-29-13-133001}.
On the local Lorentz manifold, the number of independent components of the Lagrangian form is 10; of these, six are from the curvature two-form and four are from the vierbein forms.
These degrees of freedom (d.o.f\@.) correspond to the total d.o.f\@. of the Poincar\'{e} group.
The Lagrangian form is now treated as a functional with two independent forms $(\www, \eee)$ and is independently applied a variational operation.
This method is commonly known as the Palatini method\cite{Palatini2008,ferreris}. 

Equations of motion with respect to the two independent functions (spin and vierbein forms) can be obtained by requiring a stationary condition for the variation of action for each form. 
From the variation with respect to the spin form, an equation of motion is obtained as follows:
\begin{eqnarray}
\TTT^a&=&d\eee^a+\cGR\hspace{.1em}\www^a_{~\bcdot}\wedge\eee^\bcdot=0.\label{torsion less}
\end{eqnarray}
This is referred to as the torsionless condition, which is obtained by solving the equation of motion rather than by being an independent constraint on the spacetime manifold.
It includes six independent equations, i.e., the same number of independent components as the spin form. 
Consequently, the spin form can be uniquely determined from (\ref{torsion less}) when vierbein forms are provided.
Following this, by taking variation with respect to the vierbein form allows one to obtain an equation of motion as follows:
\begin{eqnarray}
\frac{1}{2}\epsilon_{a\bcdots\bcdot}
\RRR^\bcdots\wedge\eee^\bcdot
-\Lambda_c\hspace{.1em}\VVV_a
=
\frac{1}{2}\epsilon_{a\bcdots\bcdot}
\left(
d\www^\bcdots+\cGR\www^\bcdot_{~\star}\wedge\www^{\star\bcdot}
\right)\wedge\eee^\bcdot
-\Lambda_c\hspace{.1em}\VVV_a
=0,\label{EoM2}
\end{eqnarray}
where $\VVV_a:=[\overline{\eee\wedge\eee\wedge\eee}]_a=\epsilon_{a\bcdots\bcdot}\eee^\bcdot\wedge\eee^\bcdot\wedge\eee^\bcdot/3!$ is a three-dimensional volume form.
This formulation is the Einstein equation without any matter fields, where the curvature and vierbein forms can be uniquely determined by simultaneously solving the equations of motion (\ref{torsion less}) and (\ref{EoM2}) as a quotient set of ${GL}(4,\R)$ diffeomorphism. 
Hereafter, the cosmological constant is set to zero.

\subsection{Hamiltonian formalism}\label{Hamiltonian}
The Hamiltonian formalism is constructed by identifying the fundamental forms as $(\www,\SSS)$\cite{Kurihara_2020}.
When the spin form $\www$ is identified as the general configuration variable, the canonical momentum $\MMM$ is given by
\begin{eqnarray*}
\MMM_{ab}:=\frac{\delta{\LLL_G}}{\delta\left(d\www^{ab}\right)}=\frac{1}{\kappa\hbar}\SSS_{ab}.
\end{eqnarray*}
This is consistent with the choice of phase space $(\www,\SSS)$ in the principal Poincar\'{e} bundle.
A classical form of the Hamiltonian can be obtained from the Lagrangian using the Legendre transformation as follows:
\begin{eqnarray}
\HHH_G&:=&
\frac{1}{2}\MMM_\bcdots\wedge d\www^\bcdots-{\LLL_G}=
-\frac{\cGR}{2\kappa\hbar}
\www^{\star}_{~\bcdot}\wedge\www^{\bcdot\star}\wedge\SSS_{\star\star}\hspace{.2em}.\label{CHG}
\end{eqnarray}
In our convention, the Hamiltonian form has null physical dimensions, and it is not invariant under the local ${SO}(1,3)$.
As expected, the Einstein equation and torsionless condition are obtained as canonical equations of motion.

The Poisson bracket is then provided in its covariant formalism as follows:
\begin{eqnarray*}
\left\{\aaa,\bbb\right\}_\mathrm{PB}&:=&
\frac{\delta\aaa}{\delta\www^\bcdots}\wedge\frac{\delta\bbb}{\delta\SSS_\bcdots}-
\frac{\delta\bbb}{\delta\www^\bcdots}\wedge\frac{\delta\aaa}{\delta\SSS_\bcdots},
\end{eqnarray*}
where $\aaa\in\Omega^p$ and $\bbb\in\Omega^q$ for $0\leq p,q\in\Z$.
The Poisson brackets for its fundamental forms are obtained as
\begin{eqnarray}
\begin{array}{cl}
\left\{\www^{a_1a_2},\www^{a_3a_4}\right\}_\mathrm{PB}&=
\left\{\SSS_{b_1b_2},\SSS_{b_3b_4}\right\}_\mathrm{PB}~=~0,\\
\left\{\www^{a_1a_2},\SSS_{b_1b_2}\right\}_\mathrm{PB}&=
\delta^{[a_1}_{b_1}\delta^{a_2]}_{b_2},
\end{array}
\label{pb}
\end{eqnarray}
where $\delta^{[a_1}_{b_1}\delta^{a_2]}_{b_2}:=\delta^{a_1}_{b_1}\delta^{a_2}_{b_2}-\delta^{a_2}_{b_1}\delta^{a_1}_{b_2}$.
The Hamiltonian form can be understood as the generator of the total derivative of a given form because the Poisson brackets between the fundamental and Hamiltonian forms provide 
\begin{eqnarray*}
\epsilon_{a\bcdots\bcdot}\left\{\www^\bcdots,\HHH_G\right\}_\mathrm{PB}\wedge\eee^\bcdot&=&
-\cGR\hspace{.1em}\epsilon_{a\bcdots\bcdot}\www^\bcdot_{~\star}\wedge\www^{\star\bcdot}\wedge\eee^\bcdot
=\epsilon_{a\bcdots\bcdot}d\www^\bcdots\wedge\eee^\bcdot,\\
\left\{\SSS_{ab},\HHH_G\right\}_\mathrm{PB}&=&
-\left(-\cGR\hspace{.1em}
\eta_{b\bcdot}\www^\bcdots\wedge\SSS_{\bcdot a}\right)=d\SSS_{ab},
\end{eqnarray*}
in which the canonical equations of motion are used.

\subsection{co-Poincar\'{e} symmetry and Symplectic structure}\label{symplectic structure}
It is considered to extend a Lorentz group to Poincar\'{e} group $I\hspace{-.1em}SO(1,3)=SO(1,3)\ltimes T{(4)}$.
Representations of Lie algebra $\iii\sss\ooo(1,3)$ is obtained using a trivial basis as follows:
\begin{eqnarray}
\left[P_{a},P_{b}\right]&=&0,\label{LAP1}\\
\left[J_{ab},P_{c}\right]&=&-\eta_{ac}P_{b}+\eta_{bc}P_{a},\label{LAP2}\\
\left[J_{ab},J_{cd}\right]&=&
-\eta_{ac}J_{bd}+\eta_{bc}J_{ad}
-\eta_{bd}J_{ac}+\eta_{ad}J_{bc}.\label{LAP3}
\end{eqnarray}
where $P_a$ and $J_{ab}$ are generators of $T^4$ and $SO(1,3)$ groups, respectively.
Although it is known that the Einstein--Hilbert gravitational Lagrangian does not have a Poinca\'{e} symmetry\cite{0264-9381-29-13-133001}, it has the co-Poincar\'{e} symmetry, which is introduced into the local Lorentz manifold by the author in previous works\cite{doi:10.1063/1.4990708,Kurihara_2020}. 
The co-Poincar\'{e} symmetry is the symmetry in which the translation operator  is replaced by the co-translation operator, whose generator is defined as $P_{ab}:=P_a\iota_b$, where $\iota_b$ is a contraction with respect to trivial frame field $\partial_b$.
A precise definition and some properties of co-translation operator are provided in Appendix~\ref{cotrans}.
For the co-Poincar\'{e} group, the second Lie algebra (\ref{LAP2}) is replaced by a following relation:
\begin{eqnarray}
\left[J_{ab},P_{cd}\right]&=&-\eta_{ac}P_{bd}+\eta_{bc}P_{ad}.\label{LAP2-2}
\end{eqnarray}
As a results, Lie algebra of the co-Poincar\'{e} group is provided as (\ref{LAP1}), (\ref{LAP3}) and (\ref{LAP2-2}), which corresponds to the isomorphism $SO(4)\simeq SU(2)\times SU(2)$.
This isomorphism is not unique, and there is another isomorphism, $SO(2,2)\simeq SL(2,R)\times SL(2,L)$, which gives another Lie algebra such that $\left[P_{ab},P_{cd}\right]\neq0$.
The latter case is not treated in this study.

The structure constants $\FF$ of the co-Poincar\'{e} group can be obtained from above Lie algebra through a relation\cite{fre2012gravity}, $[\Theta_I,\Theta_J]:=\FF_{~IJ}^{K}\hspace{.1em}\Theta_K$, where\footnote{
Capital Roman letters indicate indices of the structural group, and the Einstein convention is also applied on them.
}
$\left[\Theta_I\right]_{ab;cd}:=\left(\left[\Theta_{1}\right]_{ab},\left[\Theta_{2}\right]_{cd}\right)=\left(P_{ab},J_{cd}\right)$.
Lie algebra with respect to the co-Poincar\'{e} symmetry is denoted as $\ggg_{cP}$. 
Each component of $\FF$ is provided by direct calculation using trivial basis as
\begin{eqnarray}
\left\{
\begin{array}{l}
\FF_{~11}^{1}=\FF_{~11}^{2}=\FF_{~12}^{2}=
\FF_{~21}^{2}=\FF_{~22}^{1}=0,\\
\left[\FF_{~12}^{1}\right]_{ab;cd}^{ef}=
-\left[\FF_{~21}^{1}\right]_{cd;ab}^{ef}=
\eta_{ac}\delta^e_b\delta^f_d-\eta_{bc}\delta^e_a\delta^f_d,\\
\left[\FF_{~22}^{2}\right]_{ab;cd}^{ef}=
-\eta_{ac}\delta_b^e\delta_d^f+\eta_{bc}\delta_a^e\delta_d^f
-\eta_{bd}\delta_a^e\delta_c^f+\eta_{ad}\delta_b^e\delta_c^f.
\end{array}
\right.\label{LiecoP}
\end{eqnarray}

Connection form $\AAA_{\cP}$ and curvature form $\FFF_{\cP}$ are respectively introduced as Lie-algebra valued one- and two-forms with respect to co-Poincar\'{e} group, and they are expressed using a trivial basis as
\begin{eqnarray}
\AAA_{\cP}&=&\www^\bcdots\otimes J_\bcdots+
\SSS_\bcdots\otimes P^\bcdots
\hspace{2.2em}\in\Omega^1\otimes\ggg_{\cP},\label{conne2}\\
\FFF_{\cP}&=&
\RRR^\bcdots\otimes J_\bcdots+ 
d_\www\left(\SSS_\bcdots\otimes P^\bcdots\right)
\in\Omega^2\otimes\ggg_{\cP}.\label{cur2}
\end{eqnarray}

Previous work\cite{doi:10.1063/1.4990708,Kurihara_2020} has also demonstrated the Einstein--Hilbert gravitational action without the cosmological constant can be written as
\begin{eqnarray}
\LLL_{\cP}=-\frac{1}{\hbar\kappa}\Tr_{\cP}[\FFF_{\cP}\wedge\FFF_{\cP}].\label{cPL}
\end{eqnarray}
The gravitational theory proposed in this study is termed the Chern--Weil gravitational theory (CWG theory).

The Hamiltonian formalism of general relativity introduced in section \ref{Hamiltonian} is consistent with the symplectic structure.
A symplectic manifold of general relativity for a pure gravitational case has been presented as Theorem 3.2 in Ref.\cite{Kurihara_2020}.
Then, according to the standard procedure of geometrical quantization, a prequantization bundle of general relativity was constructed.

Geometrical quantization of a symplectic manifold involves two steps\cite{bates1997lectures, nair2005quantum}. The first step is to determine a polarization (which divides a phase space into half) in a symplectic manifold and to construct Hilbert space $\HH$; and this step is referred to as prequantization.
The second step is to introduce automorphism Aut$(\HH,\hat{H})$, where $\hat{H}$ in a Hamiltonian operator defined in $\HH$.
This quantization relies on the fact that the contact form has a topological class (second Chern class) that has been proven as Theorem 4.2 in Ref.\cite{doi:10.1063/1.4990708}; proofs of the theorem and related a remark are given in Appendix~\ref{cotrans}.

\subsection{Comparison between general relativity and the Yang-Mills theory}\label{YMtheory}
From a geometrical viewpoint on the Yang--Mills theory, gauge potential $\A^I_a$ and field strength $\f^I_{ab}$ correspond to a connection and curvature of the principal gauge-bundle (e.g., Ref.\cite{nair2005quantum}).
Moreover, the Dirac-spinor field $\psi$ is interpreted as a Clifford-algebra valued section over the bundle.
A gauge-potential form, which is a Lie-algebra valued one-form, is defined as follows:
\begin{eqnarray*}
\AAA_\SU:=\A^I_a\hspace{.1em}\eee^a\otimes{t_I},
\end{eqnarray*}
where ${t_I}$ is a Lie algebra of a gauge group.
The field strength (curvature) of the Yang--Mills theory in a curved spacetime manifold is defined as 
\begin{eqnarray*}
\FFF_\SU:=\FFF_\SU^I\hspace{.1em}\otimes t_I&=&d_\www\AAA_\SU
-i\hspace{.1em}{c_{\hspace{-.1em}Y\hspace{-.2em}M}}\hspace{.1em}\AAA_\SU\wedge\AAA_\SU,\nonumber\\
&=&
\left(d\AAA_\SU^I+\cGR\hspace{.1em}\www\wedge\AAA_\SU^I+
\frac{c_{\hspace{-.1em}Y\hspace{-.2em}M}}{2}f^I_{~JK}\hspace{.1em}
\AAA_\SU^J\wedge\AAA_\SU^K
\right)\otimes{t_I},\label{SUCRV}
\end{eqnarray*}
where $f^I_{~JK}$ is a structure constant of a gauge group and $c_{Y\hspace{-.2em}M}$ is a coupling constant for the Yang--Mills field.
Although the standard formalization of classical general relativity and a theory of principal bundle in mathematics do not have a coupling constant in the curvature's definition, it is natural to put it in covariant derivative $d_\www$ under the comparison between general relativity and Yang--Mills theory.
It is worth noting that constant $c_{gr}$ must have a quantum correction after quantization. 
%
The Yang--Mills curvature and the second term in parenthesis are expressed using the standard basis in $\TsM$ as 
\begin{eqnarray*}
\FFF_\SU=\frac{1}{2}\f^I_{ab}\hspace{.1em}\eee^a\hspace{-.1em}\wedge\eee^b\otimes t_I,&{\rm and}&
\www\wedge\AAA_\SU^I=\A^I_\bcdot\hspace{.1em}\www^\bcdot_{~\star}\wedge\eee^\star.
\end{eqnarray*}
In this expression, the Lagrangian for the gauge field is provided as
\begin{eqnarray}
\LLL_{\rm gauge}&:=&-\frac{1}{2}\hspace{.1em}\Tr_{\SU\oplus\TsM}\left[{\FFF}_\SU\wedge{\overline\FFF}_{\SU}\right]
\label{GFL}
\end{eqnarray}
where $\overline{\FFF}_\SU:=\eta^{\bcdot\star}\eta^{\bcdot\star}\f^I_\bcdots\SSS_\stars$.
Similarities between the co-Poincar\'{e} Lagrangian (\ref{cPL}) and gauge-field Lagrangian (\ref{GFL}) are clearly observed and is not exhibited with the standard Einstein--Hilbert Lagrangian $\LLL_{G}$ but with co-Poincar\'{e} Lagrangian $\LLL_{\cP}$.
The co-Poincar\'{e} symmetry in general relativity is essential also in this comparison.

Classical Einstein's gravitational theory can be understood as a gauge theory using the gauge group of local $SO(1,3)$\cite{PhysRev.101.1597}.
These correspondences are summarized in Table-\ref{comparison}.
The abovementioned similarity also suggests that the spin-form should be chosen as a physical field. 
	\begin{table}[t]
		\begin{center}
			\begin{tabular}{c||c|c}
Property & General Relativity & Yang-Mills Theory \\
\hline
gauge group & local $SO(1,3)$ & e.g., $SU(N)$\\
connection & spin form: $\www$ & gauge potential: $\AAA_\SU$ \\
curvature & curvature form: $\RRR$ & field strength: $\FFF_\SU$ \\
Lagrangian & 
$\Tr_{\cP}[\FFF_{\cP}\wedge\FFF_{\cP}]$ &
$\Tr_{\SU\oplus\TsM}\left[{\FFF}_\SU\wedge{\overline\FFF}_{\SU}\right]$ \\
section & vierbein form: $\eee$ & Dirac spinor: $\psi$ \\
coupling constant & $\cGR$ & $c_{Y\hspace{-.2em}M}$
			\end{tabular}
			\caption{Comparison between general relativity and the Yang--Mills theory at the classical level.}
\label{comparison}
		\end{center}
	\end{table}

\section{Canonical quantization of general relativity}\label{cqg}
The quantum field theory of general relativity under the Kugo--Ojima formalism\cite{kugo1979local, Kugo1978459} has been discussed by Nakanishi in a series of papers\cite{Nakanishi:1977gt,Nakanishi:1978zx,Nakanishi:1978np,Nakanishi:1978rt,Nakanishi:1979ff,Nakanishi:1979rr,Nakanishi:1979rq,Nakanishi:1979um,Nakanishi:1980rf,Nakanishi:1980db,Nakanishi:1980yf,Nakanishi:1980hp,Nakanishi:1980kf,Nakanishi:1981fj}
and summarized in reports\cite{nakanishi1990covariant, nakanishi1983manifestly}.
It is a self-consistent theory, beeing both physically and mathematically rigorous.
We follow their method to quantize our theory in this section.

\subsection{BRST-transformation}\label{6-1}
Dirac's procedure of canonical quantization for the system with restraints, such as the Yang--Mills gauge theory, requires the additional supplementary terms to remove unphysical d.o.f\@., e.g., gauge fixing and Faddeev--Popov ghost terms. 
Following the removal of unphysical d.o.f\@., the Lagrangian retains BRST symmetry.
A systematic approach to the quantization of such a restraint system has been established by Nakanishi\cite{Nakanishi01061966} and Kugo and Ojima\cite{kugo1979local,Kugo1978459}.
The BRST transformation of general relativity is thus introduced according to their method\cite{Nakanishi01061978, Nakanishi01071978}.

The necessary auxiliary field and Faddeev--Popov ghost and antighost fields, are introduced as follows:
\begin{itemize}
\item auxiliary field:\hspace{3em}$\beta_{\mu~b}^{~a}(x)$,
\item ghost fields:\hspace{4em}$\chi_{~b}^a(x)$ and $\chi_\mu(x)$,
\item antighost field:\hspace{2.7em}$\tilde{\chi}_{\mu~b}^{~a}(x)$.
\end{itemize}
These fields are assumed to be Hermitian functions (operators) where $\tilde{\bullet}$ represents an antighost.
The ghost field can be factorized into two parts, i.e., a local Lorentz part $\chi^a_{~b}$ and a global part $\chi_\mu$; however, the antighost is not factorized in the same way.
The $\beta$-field and (anti)ghost fields are global vectors with respect to Greek indices, they are not local Lorentz tensors with respect to Roman indices and are transformed in the same way as the spin connection as shown in (\ref{LT}) under ${\GSO}$.
Roman indices can be raised or lowered by $\eta^{ab}$ or $\eta_{ab}$, e.g., $\beta_\mu^{~ab}=\beta_{\mu~\bcdot}^{~a}\eta^{\bcdot b}$ and so on.
Ghost and antighost fields are quantized under anticommutation relations as in the standard method.
These fields are antisymmetric under an exchange between two Roman indices. 

The BRST transformation, denoted as $\delBRST[\bullet]$ in this study, satisfies four rules introduced in previous work\cite{Nakanishi01071978}. 
First, the BRST transformation of the coordinate vector in $\MM$ should obey the general linear transformation as follows:
\begin{eqnarray*}
\delBRST\left[x^\mu\right]&=&\chi^\mu.
\end{eqnarray*}
In addition, we require the postulate given in \cite{Nakanishi01071978}, such as
\begin{eqnarray*}
\delBRST\left[\partial_\mu X\right]&=&\partial_\mu\delBRST\left[X\right]
-\left(\partial_\mu\delBRST\left[x^\nu\right]\right)\partial_\nu X,
\end{eqnarray*}
where $X$ is any field defined in $\TMM$.
For differential forms on $\TsMM$, the BRST-transformation acts as
\begin{eqnarray*}
\delBRST\left[dx^\mu\right]&=&\left(\partial_\nu\delBRST\left[x^\mu\right]\right)dx^\nu
=d\left(\delBRST\left[x^\mu\right]\right)
=~d\chi^\mu.
\end{eqnarray*}
Therefore, the BRST transformation and external derivative are commute with each other, i.e., 
\begin{eqnarray}
\left[\delBRST,d\right]\bullet&=&\delBRST\left[d\bullet\right]-
d\left(\delBRST\left[\bullet\right]\right)=~0.\label{dBRST}
\end{eqnarray}
The BRST transformation of above fields is defined as follows:
\begin{eqnarray}
\left\{
\begin{array}{cl}
\delBRST\left[\beta_{\mu~b}^{~a}\right]&=~0,\\
\delBRST\left[\chi_\mu\right]&=~-
g_{\mu\rho}\left(
\partial^\rho\chi^\nu+
\partial^\nu\chi^\rho
\right)\chi_\nu,\\
\delBRST\left[\chi^{a}_{~b}\right]&=~\chi^a_{~\bcdot}\chi^\bcdot_{~b},\\
\delBRST\left[\tilde{\chi}_{\mu~b}^{~a}\right]&=~i\beta_{\mu~b}^{~a}.
\end{array}
\right.\label{BRSTchi}
\end{eqnarray}
The BRST transformation satisfies the following Leibniz rule:
\begin{eqnarray}
\delBRST\left[XY\right]&=&\delBRST\left[X\right]Y+\epsilon_XX\delBRST\left[Y\right],\label{Leib}
\end{eqnarray}
where the signature $\epsilon_X=-1$ for $X\in\{\chi_{~b}^a(x),\chi_\mu(x), \tilde{\chi}_{\mu~b}^{~a}(x)\}$, and $\epsilon_X=+1$ otherwise.
The BRST transformations of vierbein and spin connection are defined as
\begin{eqnarray*}
\delBRST\left[\Varepsilon^a_\mu\right]&=&
\Varepsilon^{~\bcdot}_\mu~\chi_{~\bcdot}^{a}
-\left(\partial_\mu\chi^{\nu}\right)\Varepsilon^a_\nu,\\
\delBRST\left[\omega^{~ab}_{\mu}\right]&=&
\omega_{\mu}^{~a\bcdot}~\chi_{~\bcdot}^{b}+
\omega_{\mu}^{~\bcdot b}~\chi_{~\bcdot}^{a}-\cGR^{-1}
\partial_\mu\chi^{a}_{~\bcdot}~\eta^{b\bcdot}
-\left(
\partial_\mu\chi^\nu
\right)\omega^{~ab}_{\nu}.
\end{eqnarray*}
The BRST transformation of the vierbein inverse $\Varepsilon^\mu_a$ is obtained after simple calculations as
\begin{eqnarray*}
\delBRST\left[\Varepsilon_a^\mu\right]&=&-\chi^{\bcdot}_{~a}~\Varepsilon_{\bcdot}^{\mu}
+\Varepsilon_a^\nu\left(\partial_\nu\chi^{\mu}\right),
\end{eqnarray*}
where $\delBRST\left[\Varepsilon^a_\mu~\Varepsilon_a^\nu\right]=\delBRST\left[\delta_\mu^\nu\right]=0$ is used.
Accordingly, the BRST transformations of the metric tensor and its inverse are obtained from above relations as
\begin{eqnarray}
\delBRST\left[g_{\mu\nu}\right]&=&\delBRST\left[
\eta_\bcdots
\eee^\bcdot_\mu
\eee^\bcdot_\nu
\right]=
-g_{\mu\rho}\partial_\nu\chi^\rho
-g_{\nu\rho}\partial_\mu\chi^\rho,\label{BRSTg}\\
\delBRST\left[g^{\mu\nu}\right]&=&\delBRST\left[
\eta^\bcdots
\eee_\bcdot^\mu
\eee_\bcdot^\nu
\right]\hspace{.2em}=\hspace{.7em}
g^{\mu\rho}\partial_\rho\chi^\nu+
g^{\rho\nu}\partial_\rho\chi^\mu.\label{BRSTgi}
\end{eqnarray}
Here, we use antisymmetry of the ghost filed $\chi_{ab}=-\chi_{ba}$ and symmetry of the metric tensor. 
The BRST transformations of the vierbein and spin forms are provided from above results as
\begin{eqnarray*}
\delBRST\left[\eee^a\right]&=&
\chi_{~\bcdot}^{a}\eee^\bcdot,\\
\delBRST\left[\www^{ab}\right]
&=&\www^{a\hspace{.1em}\bcdot}\chi_{~\bcdot}^{b}
+\www^{\bcdot\hspace{.1em} b}~\chi_{~\bcdot}^{a}
-{\cGR}^{-1}d\chi^{a}_{~\bcdot}~\eta^{b\hspace{.1em}\bcdot}.
\end{eqnarray*}

The one-forms corresponding to the above quantities are introduced as
\begin{eqnarray*}
\bbb^{ab}&=&\beta_{\mu}^{~ab}\hspace{.1em}\partial_\nu\chi^\mu dx^\nu=
\beta_{\mu}^{~ab}~d\chi^\mu,\\
\ccc^{a}&=&\chi^a_{~\bcdot}\hspace{.1em}\eee^\bcdot\hspace{3.3em}=
\chi^a_{~\bcdot}\hspace{.1em}\Varepsilon^\bcdot_\mu\hspace{.1em}dx^\mu,\\
\tilde{\ccc}^{ab}&=&\tilde{\chi}_{\mu}^{~ab}\partial_\nu\chi^\mu\hspace{.1em}dx^\nu=
\tilde{\chi}_{\mu}^{~ab}\hspace{.1em}d\chi^\mu,
\end{eqnarray*}
which are referred to as b-, ghost-, and antighost-forms, respectively.
We note that $\bbb^{ab}$ and $\ccc^{a}$ are anticommutable, and $\tilde{\ccc}^{ab}$ is commutable.
The BRST transformations of those forms are provided as
\begin{eqnarray*}
\delBRST\left[\bbb^{ab}\right]=\delBRST\left[\ccc^a\right]=0&{\rm and}&
\delBRST\left[\tilde{\ccc}^{ab}\right]=i\bbb^{ab},
\end{eqnarray*}
and that of the vierbein form is provided as
\begin{eqnarray*}
\delBRST\left[\eee^a\right]&=&\ccc^a.
\end{eqnarray*}
By direct calculations, it is confirmed that the BRST transformation is nilpotent for all forms and fields defined above.
A proof of nilpotency for all necessary forms is provided in Appendix~\ref{ap1}.

\subsection{Quantum Lagrangian for gravity}\label{6-2}
We start from the Lagrangian of classical gravity presented in equation (\ref{Lagrangian44}) with $\Lambda_c=0$.
The quantum Lagrangian for gravity is completed by adding gauge-fixing and Faddeev--Popov ghost terms as follows:
\begin{eqnarray}
\LLL_{Q\hspace{-.1em}G}&=&\LLL_G+\LLLGF+\LLLFP,\label{QLGF}
\end{eqnarray}
where $\LLLGF$ and $\LLLFP$ are referred to as the gauge-fixing Lagrangian and the Faddeev--Popov Lagrangian forms, respectively; their explicit forms will be given later in this section.
The quantum Lagrangian form is constructed to maintain invariance and nilpotency under the BRST transformations that the Lagrangian form is a BRST-null object, i.e., $\delBRST\left[\LLLQG\right]=0$.

The gauge-fixing Lagrangian form is determined to obtain a desired gauge-fixing condition. 
Here, we require the de\hspace{.2em}Donder condition for the gauge fixing, which is represented using the metric tensor as $\partial_\mu\left(\sqrt{-g}~g^{\mu\nu}\right)=0$.
An alternative representation is also possible as $\Gamma^\lambda_{~\mu\nu}~g^{\mu\nu}=0$.
The corresponding representation in the vierbein formalism is expressed as $d\SSS_{ab}=0$.

The simplest candidate for the gauge-fixing form, which provides the de\hspace{.2em}Donder condition as an Euler--Lagrange equation with respect to the variation of the $b$-form, is as follows:
\begin{eqnarray}
\LLLGF&=&
-\frac{1}{2\hbar\kappa}\left(d\bbb^\bcdots+\alpha\bbb^\bcdot_{~\star}\wedge\bbb^{\star\bcdot}
\right)\wedge\SSS_\bcdots,\label{LGF}
\end{eqnarray} 
where $\alpha\in\R$ is a gauge fixing parameter.
In a practical sense, the $\alpha$-term vanishes autonomously, i.e., $\bbb^\bcdot_{~\star}\wedge\bbb^{\star\bcdot}\wedge\SSS_\bcdots=0$, owing to the anticommutation of the $b$-form.
Herein, it is retained for purpose of discussing the function of the gauge fixing term.
Terms $d\bbb^{ab}+\alpha\bbb^a_{~\bcdot}\wedge\bbb^{\bcdot b}$ (without multiplying the surface form) retain their Lorentz invariance when $\alpha=\cGR$ because it has the same structure as the curvature form when using the $b$-form instead of the spin form.
The gauge-fixing Lagrangian must be as follows:
\begin{eqnarray*}
\LLLGF&=&\frac{i}{2\hbar\kappa}
\delBRST\left[d\tilde{\ccc}^\bcdots+\alpha
\tilde{\ccc}^{\bcdot}_{~\star}\wedge\bbb^{\star\bcdot}
\right]\wedge\SSS_\bcdots\hspace{.2em}.
\end{eqnarray*}
Therefore, if the Faddeev--Popov form is chosen as
\begin{eqnarray*}
\LLLFP&=&-\frac{i}{2\hbar\kappa}\left(
d\tilde{\ccc}^\bcdots+
\alpha
\tilde{\ccc}^{\bcdot}_{~\star}\wedge\bbb^{\star\bcdot}
\right)\wedge\delBRST\left[\SSS_\bcdots\right],
\end{eqnarray*} 
the sum of the gauge-fixing and Faddeev--Popov terms is BRST-null because both the antighost and surface forms are nilpotent.
Accordingly, the Faddeev--Popov Lagrangian-form can be obtained as
\begin{eqnarray}
\LLLFP&=&
-\frac{i}{2\hbar\kappa}\left(d\tilde{\ccc}^\bcdots+
\alpha
\tilde{\ccc}^{\bcdot}_{~\star}\wedge\bbb^{\star\bcdot}\right)
\wedge\left(\epsilon_{\bcdots c_1c_2}\chi^{c_1}_{~c_3}
\eee^{c_3}\wedge\eee^{c_2}
\right),\nonumber\\
&=&-\frac{i}{2\hbar\kappa}\left(
d\tilde{\ccc}^\bcdots+
\alpha
\tilde{\ccc}^{\bcdot}_{~\star}\wedge\bbb^{\star\bcdot}\right)
\wedge\ccc^\star\wedge\overline{\eee}_{\star\bcdots}.\label{LFP}
\end{eqnarray}
The Faddeev--Popov Lagrangian-form is not Lorentz invariant, even if $\alpha={\cGR}$, and the $\alpha$-term in $\LLLFP$ does not vanish after multiplying with $\delBRST\left[\SSS_{ab}\right]$.
It is thus confirmed that the Faddeev--Popov Lagrangian form validates the Lorentz invariance. 
Finally, a quantum Lagrangian with any value of $\alpha$ satisfies the abovementioned requirements.
According to particle physics terminologies, this is referred to as the Landau gauge when $\alpha=0$, and Feynman-gauge when $\alpha=1$.

An Euler--Lagrange equation of motion can be obtained under the variational principle. 
All relevant equations of motion are listed in Appendix~\ref{app2}. 

\subsection{Commutation relations}
From the quantum Lagrangian (\ref{QLGF}), the formal expressions of the commutation relations between forms at two points $x,y\in\MM$ can be selected as follows:
\begin{eqnarray}
\left[\widehat{\www}^{ab}(x),\widehat{\SSS}_{cd}(y)\right]&=&-i\delta^{(4)}(x-y)
\delta^{[a}_{c}\delta^{b]}_{d},\label{CR1}\\
\left[\widehat{\bbb}^{ab}(x),\widehat{\SSS}_{cd}(y)\right]&=&-i\delta^{(4)}(x-y)
\delta^{[a}_{c}\delta^{b]}_{d},\label{CR2}\\
\left\{\widehat{\tilde\ccc}^{ab}(x),(\widehat{\ccc^\bcdot\wedge\overline{\eee}}_{\bcdot cd})(y)\right\}&=&-i\delta^{(4)}(x-y)
\delta^{[a}_{c}\delta^{b]}_{d}.\label{CR3}
\end{eqnarray} 
where $\widehat{\bullet}$ denotes as the operator corresponding to the field or form $\bullet$.
Note that $\delta\LLLQG/\delta\bbb^{ab}=\SSS_{ab}$, and the form $\ccc^a$ are not necessarily treated as canonical variables.
In our discussion, the abovementioned commutation relations are rewritten using the original fields instead of forms as follows. 
First, the commutation relations (\ref{CR1}) and (\ref{CR2}) are given by
\begin{eqnarray}
\left[\widehat{\omega}_{\mu~\bcdot}^{~a}(x)\eta^{\bcdot b},\widehat{\SSS}_{cd}(y)\right]dx^\mu&=&-i\delta^{(3)}(x-y)
\delta^{[a}_{c}\delta^{b]}_{d},~~\label{fCR1}\\
\left[\widehat{\beta}_{\nu~\bcdot}^{~a}(x)\eta^{\bcdot b}\partial_\mu\widehat{\chi}^\nu,
\widehat{\SSS}_{cd}(y)\right]dx^\mu&=&-i\delta^{(3)}(x-y)
\delta^{[a}_{c}\delta^{b]}_{d}.~~\label{fCR2}
\end{eqnarray}
These are still formal expressions that omit a two-dimensional integration measure in the right-hand side to avoid long expressions.
We use these formal expressions because they are sufficient for the following discussion.
To obtain the field-representation for (\ref{CR3}), some manipulation is required.
The left had side of (\ref{CR3}) can be expressed as
\begin{eqnarray*}
(\ref{CR3})&=&
\epsilon_{cde_1e_2}~\widehat{\chi}^{e_1}_{~~\bcdot}~\epsilon^{\bcdot e_2 f_1f_2}\left\{
\widehat{\tilde{\chi}}_{\nu~\star}^{~a}\eta^{\star b}\partial_\mu\widehat{\chi}^\nu,\widehat{\SSS}_{f_1f_2}
\right\}dx^\mu.
\end{eqnarray*}
Let us introduce matrix expressions as
\begin{eqnarray*}
\widehat{\bm{\epsilon\chi\bar{\epsilon}}}&:=&
\epsilon_{ab e_1e_2}~\widehat{\chi}^{e_2}_{~~\bcdot}~\epsilon^{\bcdot e_1cd},\\
\widehat{\{\tilde{\bm\chi},\bm\SSS\}}&:=&
\left\{
\widehat{\tilde{\chi}}_{\nu~\bcdot}^{~a}\eta^{\bcdot b}\partial_\mu\widehat{\chi}^\nu
,\widehat{\SSS}_{cd}
\right\}dx^\mu,
\end{eqnarray*}
where matrices $\widehat{\bm{\epsilon\chi\bar{\epsilon}}}$ and $\widehat{\{\tilde{\bm\chi},\bm\SSS\}}$ are $4\times4$ matrices, whose elements are also $4\times4$ matrices.
The commutation relation (\ref{CR3}) can then be expressed as 
\begin{eqnarray}
\widehat{\bm{\epsilon\chi\bar{\epsilon}}}\hspace{.1em}\bcdot\hspace{.1em}
\widehat{\{\tilde{\bm\chi},\bm\SSS\}}&=&
-i\delta^{(3)}(x-y)~\bm{\mathrm I},\label{CR3-2}
\end{eqnarray}
where $\bm{\mathrm I}$ is the matrix expression of $\delta^{[a}_c\delta^{b]}_d$, which is also a $4\times4$ matrix of $4\times4$ matrices.
If each element of matrix $\widehat{\bm{\epsilon\chi\bar{\epsilon}}}$ is invertible, one can obtain the commutation relation of $\widehat{\{\tilde{\bm\chi},\bm\SSS\}}$ by applying the inverse matrix in (\ref{CR3-2}).
The matrix is invertible with the exception of its the diagonal part.
The diagonal part is characterized by $\left[\widehat{\bm{\epsilon\chi\bar{\epsilon}}}\right]_{aa}=\bm{0}$, where $\bm{0}$ is a $4\times4$ zero-matrix, and it is not necessarily for the inversion because the diagonal part of $\widehat{\{\tilde{\bm\chi},\bm\SSS\}}$ is also zero-matrix owing to the antisymmetric nature of the surface form.
Therefore, a matrix inverse can be obtained as
\begin{eqnarray*}
\left[\widehat{\bm{\epsilon\chi\bar{\epsilon}}}^{-1}\right]_{IJ}&=&
\left\{
\begin{array}{cl}
\left(\left[\widehat{\bm{\epsilon\chi\bar{\epsilon}}}\right]_{JI}\right)^{-1}&(I\neq J)
\\
\bm{0}&(I=J),
\end{array}
\right.
\end{eqnarray*}
where $I$ and $J$ are the indices for each $4\times4$ matrix.
Thus, the commutation relation is given by
\begin{eqnarray}
\widehat{\{\tilde{\bm\chi},\bm\SSS\}}&=&
-i\delta^{(3)}(x-y)
\widehat{\bm{\epsilon\chi\bar{\epsilon}}}^{-1}, \label{fCR3}
\end{eqnarray}
using a matrix representation.
Each element of the matrix $\left[\widehat{\bm{\epsilon\chi\bar{\epsilon}}}^{-1}\right]_{IJ}$ is either $\pm\left(\chi^a_{~b}\right)^{-1}$ or zero.

In summary, three nonzero commutation-relations (\ref{fCR1}), (\ref{fCR2}), and (\ref{fCR3}), are obtained. 
All other commutation relations are zero.

\subsection{BRST charge}\label{BRSTCharge}
Whereas a local gauge invariance of the Lagrangian is violated owing to the gauge-fixing term, a global symmetry, which is referred to as the BRST symmetry, is remaining; and thus, the conserved Noether charge exists owing to the Noether theorem.
The Noether charge associated with a global BRST symmetry is provided using a method given in \cite{Kurihara2018} such that
\begin{eqnarray*}
\QQQ_\xi&=&\frac{1}{2}\left(
\iota_\xi\www^\bcdots
\right)\SSS_\bcdots,
\end{eqnarray*}
where $\xi^\mu$ is any vector in $T\MM$ and $\iota_\xi\aaa$ is a contraction operator between form $\aaa$ and vector $\xi$.
First, by taking the auxiliary field as the vector field $\xi^\mu$, we obtain a conserved charge as
\begin{eqnarray*}
\QQQ_\bbb&:=&
\frac{1}{2}\left(\iota_{\bbb}\www^\bcdots\right)\SSS_\bcdots=
\frac{1}{2}g^{\mu_1\mu_2}{\beta}^{~a_1}_{\mu_3~a_2}\partial_{\mu_1}\chi^{\mu_3}\omega_{\mu_2}^{~a_2a_3}~\SSS_{a_3a_1},
\end{eqnarray*}
and its BRST transformation is
\begin{eqnarray*}
\widehat{\QQQ}_\bbb&:=&\delBRST\left[\QQQ_\bbb\right]=
\frac{1}{2}{\beta}^{~a_1}_{\mu_3~a_2}\delBRST\left[g^{\mu_1\mu_2}\partial_{\mu_1}\chi^{\mu_3}\omega_{\mu_2}^{~a_2a_3}~\SSS_{a_3a_1}\right].
\end{eqnarray*}
All the fields in above BRST transformation are nilpotent, and thus, $\widehat{\QQQ}_\beta$ is BRST-null according to the remark in Appendix \ref{ap1}.
Next, when the antighost form is taken as the vector field $\xi^\mu$, the conserved charge can be obtained as
\begin{eqnarray*}
\QQQ_{\tilde\ccc}&:=&\frac{1}{2}\left(\iota_{\tilde\ccc}\www^\bcdots\right)\SSS_\bcdots=
\frac{1}{2}g^{\mu_1\mu_2}{\tilde\chi}^{~a_1}_{\mu_3~a_2}\partial_{\mu_1}\chi^{\mu_3}\omega_{\mu_2}^{~a_2a_3}~\SSS_{a_3a_1}.
\end{eqnarray*}
The BRST transformation of this charge is obtained as
\begin{eqnarray*}
\widehat{\QQQ}_{\tilde\ccc}&:=&\frac{i}{2}
\delBRST\left[\QQQ_{\tilde\ccc}\right],\\&=&
\frac{i}{4}{\tilde\chi}^{~a_1}_{\mu_1~a_2}\delBRST\left[g^{\mu_1\mu_2}\partial_{\mu_1}\chi^{\mu_3}\omega_{\mu_2}^{~a_2a_3}~\SSS_{a_3a_1}\right]
-\frac{1}{4}{\beta}^{~a_1}_{\mu_1~a_2}\left(g^{\mu_1\mu_2}\partial_{\mu_1}\chi^{\mu_3}\omega_{\mu_2}^{~a_2a_3}~\SSS_{a_3a_1}\right),
\end{eqnarray*}
where the factor $i/2$ is just a convention.
Again by taking the BRST-transformation of $\widehat{\QQQ}_{\tilde\ccc}$, one can obtain
\begin{eqnarray*}
\delBRST\left[\widehat{\QQQ}_{\tilde\ccc}\right]&=&-\frac{1}{2}
{\beta}^{~a_1}_{\mu_1~a_2}\delBRST\left[g^{\mu_1\mu_2}\partial_{\mu_1}\chi^{\mu_3}\omega_{\mu_2}^{~a_2a_3}~\SSS_{a_3a_1}\right]=
-\widehat{\QQQ}_\bbb,
\end{eqnarray*}
where the nilpotency of $(g\omega\SSS)$ is used.
The charge $\QQQ_{\tilde\ccc}$ is not nilpotent, but $\widehat{\QQQ}_{\tilde\ccc}$ is. 

Because the charge $\QQQ_\xi$ is conserved\cite{Kurihara2018}, and the  external derivative and BRST transformation are commute with each other as shown in (\ref{dBRST}), both charges are conserved as
\begin{eqnarray*}
d\widehat{\QQQ}_\xi&=&d\left(\delBRST\left[\QQQ_\xi\right]\right)=~\delBRST\left[d\QQQ_\xi\right]=~0,
\end{eqnarray*}
where $\xi=\tilde{\ccc},\bbb$.
In conclusion, the two conserved charges $\widehat{\QQQ}_\bbb$ and $\widehat{\QQQ}_{\tilde\ccc}$ are obtained such that
\begin{eqnarray*}
\widehat{\QQQ}_\bbb&=&\frac{1}{2}\beta^{~\bcdot}_{\mu~\star}~\delBRST
\left[\widehat{\QQQ}_{~~\bcdot}^{\mu\star}\right],\\
\widehat{\QQQ}_{\tilde\ccc}&=&\frac{i}{4}\tilde\chi^{~\bcdot}_{\mu~\star}~\delBRST
\left[\widehat{\QQQ}_{~~\bcdot}^{\mu\star}\right]
-\frac{1}{4}\beta^{~\bcdot}_{\mu~\star}~\widehat{\QQQ}_{~~\bcdot}^{\mu\star},\\
\end{eqnarray*}
where 
\begin{eqnarray}
\left[\widehat{\bm\QQQ}\right]_{~~b}^{\mu a}=
\widehat{\QQQ}_{~~b}^{\mu a}:=g^{\nu_1\nu_2}\partial_{\nu_1}\chi^{\mu}\omega_{\nu_2}^{~ac}~\SSS_{cb}.
\label{defQQQ}
\end{eqnarray}
These conserved charges satisfy
\begin{eqnarray*}
d\widehat{\QQQ}_\bbb=d\widehat{\QQQ}_{\tilde\ccc}=0,\hspace{1em}
\delBRST\left[\widehat{\QQQ}_{\tilde\ccc}\right]=-\widehat{\QQQ}_\bbb,&{\rm and}&
\delBRST\left[\widehat{\QQQ}_{\bbb}\right]=0.
\end{eqnarray*}
Moreover, the charge $\widehat{\QQQ}_{\bbb}$ is a generator of the BRST transformation for operators, such as
\begin{eqnarray}
i\lambda\hspace{.1em}\delBRST\left[\widehat\Omega\right]&=&\left[\widehat{\Omega},
\lambda\hspace{.1em}\widehat{\QQQ}_\bbb\right],\label{BRSTgen0}
\end{eqnarray}
where $\widehat\Omega$ is any operator and $\lambda=i(-1)^{\pm\widehat{\QQQ}_{\tilde\ccc}}$\cite{Nakanishi:1977gt}.
Especially by taking $\widehat\Omega=\widehat{\QQQ}_{\tilde\bbb}$ or $\widehat{\QQQ}_{\tilde\ccc}$, one can obtain
\begin{eqnarray}
\widehat{\QQQ}_{\bbb}^2\hspace{1em}&=&0,\label{BRSTgen1}\\
\left[\widehat{\QQQ}_{\tilde\ccc},\widehat{\QQQ}_\bbb\right]&=&-i\widehat{\QQQ}_\bbb.\label{BRSTgen2}
\end{eqnarray}
A relation $(\ref{BRSTgen1})$ immediately follows from the nilpotency of $\widehat{\QQQ}_{\bbb}$ and $\lambda=-1$ in (\ref{BRSTgen0}).
A proof of the relation (\ref{BRSTgen2}) is given in Appendix \ref{appC}.

\section{Discussion}\label{discussion}
In the previous section, the quantization of general relativity was formally performed using the Nakanishi--Kugo--Ojima method in the Heisenberg picture.
After quantization, the quantum Lagrangian (\ref{QLGF}) comprises the field operators of both physical fields (e.g., the spin-connection and vierbein-fields) and unphysical fields (e.g., auxiliary $\beta$-field and Faddeev--Popov ghost and antighost fields).
In this section, functional spaces in which these operators are defined are discussed.

Although the Lagrangian includes unphysical fields, the probabilities that such fields are observed must be zero, and the corresponding $S$-matrix must be a unitary matrix to be consistent with the probabilistic interpretation.
The Kugo--Ojima theorem used to ensure the unitarity of a physical $S$-matrix. 
Futhermore, the renormalizability of quantum general relativity and comparison with other geometrical theories are discussed.

\subsection{Hilbert space and physical states}
The functional space of the spin connection $\omega_{\mu}^{~ab}$ is formally considered as follows.
Classically, a spin-connection form is obtained as a solution of the Einstein equation, which is a nonlinear first-order differential equation.
It is a homogeneous equation for a pure gravitational equation without any gauge or matter fields.
When a classical solution (e.g., $\www^{(c)}$) is obtained, the Lorentz-transformed form ${\GSO}(\www^{(c)})$ is also the solution of the equation, and  a quotient set with respect to ${\GSO}$, $\varpi:=\widetilde\varpi/{\GSO}(\widetilde\varpi)$) is introduced in which $\widetilde\varpi$ is a set of solutions of the classical Einstein equation. 
When the local Lorentz manifold $\M$ has a Euclidean metric, $\varpi$ has the Sobolev norm $L^p_k(\www^{(c)})$, which is defined as
\begin{eqnarray*}
\left\|\www^{(c)}\right\|_{L^p_k}:=\left(
\sum_{0\leq i_1+\cdots+i_n\leq k}\int_{\Sigma_\M}
\left| 
{\partial_{i_1}\cdots\partial_{i_n}} \omega^{(c)}
\right|^p\vvv
\right)^{1/p},
\end{eqnarray*}
where $\Sigma_\M$ is a compact subset of $\M$ and $\vvv$ is a $GL(4,\R)$ invariant volume form defined in (\ref{vvv}).
The summation is performed over all possible combinations $(i_1,\cdots,i_n)$ that satisfy the condition $0\leq i_1+\cdots+i_n\leq k$, where $0\leq i_j\in\Z$.  
In our case, the dimension of manifold $\M$ is $n\hspace{-.2em}=\hspace{-.2em}4$.
The completion of $\varpi$ with respect to the Sobolev norm forms the Sobolev space, denoted as $L^p_k(\varpi)$.
The Einstein equation, however, is not elliptic even in Euclidean space, because the classical equation is invariant under the infinite groups $GL(4,\R)$ and $\SO(4)$.
However, the quantum Lagrangian, which includes (local) gauge-fixing terms, can be expected to be locally elliptic; thus, functional analysis can be applied locally around a gauge-fixing point.
In this case, $L^2_k(\varpi)$ is a Hilbert space with respect to an inner product defined as
\begin{eqnarray*}
\left(\www^{(c)},{\www^{(c)}}'\right)_{L^2_k(\varpi)}:=
\sum_{0\leq i_1+\cdots+i_n\leq k}\int_{\Sigma}
\left(\partial_{i_1}\cdots\partial_{i_n}\omega^{(c)}
\right)\cdot
\left(\partial_{i_1}\cdots\partial_{i_n}{\omega^{(c)}}'
\right)\vvv.
\end{eqnarray*}
A dot-product is defined using a metric tensor in functional space.
Hilbert space $L^2_0(\varpi)$ is denoted as $\HH^{(c)}$.

Functional space $\varpi$ is a subset of $\HH^{(c)}$ as $\varpi\subset\HH^{(c)}\subset L^2$.
The bar-dual space of $\varpi$, denoted as $\overline{\varpi}$, is simply referred to as the dual space of $\varpi$.
Dual space $\overline{\varpi}$ is also square-integrable; thus, the Gel'fand triple becomes  $\varpi\subseteq\HH^{(c)}\subseteq\overline{\varpi}\subset L^2$.
The coefficient functions for the spin form $\www^{ab}$ and its dual form $\overline\www_{ab}$ are  $\omega^{~ab}_{\mu}\in\varpi$ and $\epsilon_{ab\bcdots}\hspace{.2em}\omega^{~\bcdots}_{\mu}/2\in\overline\varpi$, respectively.
A standard bilinear form is defined as
\begin{eqnarray*}
\langle \overline\www|\www\rangle_{\HH^{(c)}}
&:=&\frac{1}{2}\overline\www_\bcdots\wedge\www^\bcdots=
\frac{1}{4}\epsilon_{\bcdott}\hspace{.2em}
\omega^{~~\bcdots}_{\mu_1}\hspace{.2em}\omega^{~~\bcdots}_{\mu_2}\hspace{.2em}
dx^{\mu_1}\wedge dx^{\mu_2}.
\end{eqnarray*}
A norm of the spin form $\|\www\|_{\HH^{(c)}}$ can be defined using the bilinear form as
\begin{eqnarray}
\left(\|\www\|_{\HH^{(c)}}\right)^{2}:=\int_{\Sigma_2}\langle\overline\www|\www\rangle_{\HH^{(c)}}
\in\R,\label{blf}
\end{eqnarray}
where ${\Sigma_2}$ is the appropriate compact two-dimensional submanifold ${\Sigma_2}\subset\M$.
An element of the dual space $\overline{\varpi}$ is provided by a linear combination of spin forms $\omega^{~ab}_{\mu}\in\varpi$; thus, functional spaces $\varpi$ and $\overline{\varpi}$ are linearly equivalent.
Consequently, the Hilbert space can be obtained as $\HH^{(c)}\simeq\varpi$, and the Gel'fand triple becomes  $\varpi\simeq\overline{\varpi}\simeq\HH^{(c)}$.
The same construction is possible for the surface form and the standard bilinear form is defined as
\begin{eqnarray*}
\langle \overline\SSS|\SSS\rangle_{\HH^{(c)}}
&:=&\frac{1}{2}\overline\SSS_\bcdots\wedge\SSS^\bcdots=
\left(3!\right)\hspace{.1em}\vvv,
\end{eqnarray*}
and so on.

For a further discussion, the following simple $(1\hspace{-.1em}+\hspace{-.1em}3)$ coordinate-decomposition is considered according to the ADM formalism\cite{PhysRev.116.1322}.
Assuming that the global spacetime manifold is filled with a congruence of geodesics whose tangent vector is time-like at any point on the line, a coordinate $x^0$ may be obtained along the time-like vector on these geodesics, and the other three coordinates are taken as the orthonormal base. 
Using this coordinate system, the three-dimensional boundary is obtained as a manifold at $x^0=\tau$ (an {\it equal-time} boundary) and the state is written as $|\Psi_\tau\rangle$.
The commutation relations corresponding to Poisson brackets (\ref{pb}) are obtained as
\begin{eqnarray}
\left[\widehat{\www}^{ab},\widehat{\www}^{cd}\right]&=&
\left[\widehat{\SSS}_{ab},\widehat{\SSS}_{cd}\right]~=~0\label{cr00}
\end{eqnarray}
and (\ref{CR1}).
The abovementioned expressions for the commutation relations are formal.
In particular, the representation of (\ref{CR1}) must be understood as follows:
\begin{eqnarray*}
\left[\widehat{\www}^{ab}(x),\widehat{\SSS}_{cd}(y)\right]&:=&
\left[\widehat{\omega}_{\mu}^{~~ab}(x),\widehat{\Varepsilon_{\nu}^{c}\Varepsilon_{\rho}^{d}}(y)\right],\\
&=&-i\delta^{[a}_{c}\delta^{b]}_{d}
\epsilon_{\mu\nu\rho\sigma}
\delta^{(3)}(x^{\mu,\nu,\rho}-y^{\mu,\nu,\rho})
\int\delta^{(1)}(x^{\sigma}-y^{\sigma})dx^{\sigma}.
\end{eqnarray*}
By setting $dx^\sigma=d\tau$ (``time'' coordinate) instead of taking a sum and performing integration with respect to $d\tau$, one can obtain the standard ``equal-time'' commutation relation.

The operator can be obtained as
\begin{eqnarray}
\widehat{\www}^{ab}:=\www^{ab},&~&
\widehat{\SSS}_{ab}:=i\frac{\delta~}{\delta\www^{ab}},\label{crqop}
\end{eqnarray}
reproducing the commutation relations (\ref{CR1}) and (\ref{cr00}).
A Schr\"{o}dinger equation becomes
\begin{eqnarray}
\widehat{\HHH}_G|\Psi_\tau\rangle&=&E_G|\Psi_\tau\rangle,
\end{eqnarray}
where $\widehat{\HHH}_G$ is the Hamiltonian operator corresponding to (\ref{CHG}) and $E_G$ is the energy eigenvalue, $|\Psi\rangle\in\HH^{(q)}$, where $\HH^{(q)}$ is a quantum Hilbert space corresponding to $\HH^{(c)}$.

Functional space $\HH^{(q)}$ is a dual space of $\HH^{(c)}$ with respect to the functional derivative operator (\ref{crqop}); it thus yields the Gel'fand triple $\HH^{(c)}\subset L^2(\TM)\subset \HH^{(c)*}=\HH^{(q)}$, where dual space $\HH^{(q)}$ is a space of complex functions.
Although the spin connection is a square-integrable function, state vectors are not necessarily holomorphic functions.
Therefore, the analyticity of state vectors must be confirmed in each case.

Because spin connection $\omega^{~a}_{\mu~c}(x)$ is chosen as a general coordinate in the symmplectic manifold of general relativity, a state vector can be considered as a functional of spin connection\cite{Witten198846}, and the norm of state vector $\Psi_\tau$ can be defined using (\ref{blf}).
Although the existence of the norm is ensured, its negative norm states $\langle\Psi_\tau | \Psi_\tau \rangle<0$ are included in $\HH^{(q)}$, which corresponds to a negative energy state and can be eliminated from the physical state.

\subsection{Kugo--Ojima theorem for the CWG theory}
Abovementioned negative-energy states can be removed form the theory according to the Kugo--Ojima theorem as follows.
The expectation value of the operator $\widehat{\cal{O}}$  can be represented as follows.
We note that the operator $\widehat{{\cal O}}(\widehat{\www},\widehat{\SSS})$ cannot be uniquely determined from the classical function ${\cal O}\left(\www,\SSS\right)$ because of the noncommutativity of quantum operators.
This issue is known as the operator ordering problem.
When the operator ordering is fixed, an expected value of a given operator is obtained;
\begin{eqnarray*}
\overline{{\cal O}}&=&
\langle\Psi_\tau|\widehat{{\cal O}}\left(\widehat{\www},\widehat{\SSS}\right)|\Psi_\tau\rangle
=\langle\Psi_\tau|\int{\cal O}\left(\www,\SSS\right)|\Psi_\tau\rangle=\int{\cal O}\left(\www,\SSS\right).
\end{eqnarray*}
For example, the Hamiltonian operator has an expected value that can be written as $E_G=\int_{\partial\Sigma} \iota_{0}\HHH_G$, which reflects the total energy in the three-dimensional space of the Universe at a given time.
A trivial ground state of $\www^{ab}=0$ exists with an eigenvalue of $E_G=0$.
Using the torsionless condition, one can obtain $d\eee^a=0$, which means that the ground state corresponds to the flat Lorentz spacetime.
This is one of the general results for CWG theory. 
It is known that the critical point of the Chern--Weil action provides a flat connection.

Owing to the existence of BRST generators of (\ref{BRSTgen1}), (\ref{BRSTgen2}) and the nilpotent nature of the BRST operator, its physical state can be defined according to the Nakanishi--Kugo--Ojima formalism as
\begin{eqnarray}
\widehat{\QQQ}_\bbb|\mathrm{phys}\rangle&=&\widehat{\QQQ}_{\tilde\ccc}|\mathrm{phys}\rangle=~0,
\end{eqnarray}
where $|\mathrm{phys}\rangle\in{\cal H}_{\mathrm{phys}}\subset{\cal H}$.
Thus, $\langle\mathrm{phys}|\mathrm{phys}\rangle\geq0$ is maintained according to the Kugo--Ojima theorem.
A transition operator $\widehat{{\mathcal S}}$ is introduced so that the matrix element
\begin{eqnarray}
p\left(\tau_2:\tau_1\right)&=&\langle\Psi_{\tau_2}|\widehat{{\mathcal S}}|\Psi_{\tau_1}\rangle
\end{eqnarray}
gives the transition probability from the state $|\Psi_{\tau_1}\rangle$ at $x^0=\tau_1$ to another state $|\Psi_{\tau_2}\rangle$ at $x^0=\tau_2$, where $|\Psi_{\tau_{1,2}}\rangle\in{\cal H}_{\mathrm{phys}}$ and $\widehat{{\mathcal S}}$ is a functional constructed using the spin, surface, auxiliary, and Faddeev-Popov forms.
Again the Kugo--Ojima theorem ensures the unitarity of this transition matrix\cite{nakanishi1990covariant}.
In conclusion, based on above discussion, it is proven that the time evolution of the Universe maintains unitarity in terms of the state vector of quantum gravity.

\subsection{Renormalizability}\label{renorm}
Thus far, renormalizability has not been discussed because the defined quantization is performed nonperturbatively.
In reality, for pure gravitational theory, including neither matter fields nor cosmological constant, several facts are known by direct calculations.
't\hspace{.2em}Hooft and Veltman confirmed that the on-shell $S$-matrix element of quantum general relativity is finite in the one-loop state\cite{'tHooft:1974bx}; however, the theory is nonrenormalisable if an interaction with scalar particles is included.
However, Berends and Gastmans reported that the amplitude of graviton-graviton scattering is not unitary, even at the tree-level\cite{BERENDS197599}.
Beyond the one-loop level, Goroff and Sagnotti have reported that the theory is nonrenormalisable at the two-loop level\cite{Goroff198581,GOROFF1986709}.

Although, the unitarity violation reported in \cite{BERENDS197599} appears to contradict our results, this is not the case. 
First, a metric tensor or vierbein is used as the standard variable of the perturbation expansion.
However, the appropriate phase-space variables for CWG action in four dimensions are the spin and surface forms.
Standard perturbation gravity using an improper expansion variable.
Moreover, the covariant perturbation gravity, which has been developed by de\hspace{.1em}Witt\cite{DeWitt:1967ub,DeWitt:1967uc} and used in studies to show nonrenormalizability\cite{'tHooft:1974bx,BERENDS197599,Goroff198581,GOROFF1986709}, uses the background field method in which the metric tensor is separated into the classical background metric $g_{\mu\nu}$ and quantum field $h_{\mu\nu}$ such as $g_{\mu\nu}\rightarrow g_{\mu\nu}+h_{\mu\nu}$.
On the other hand, the ground state of the quantum Hamiltonian must have a zero-point energy because of the quantum effect.
Although quantum general relativity in the covariant perturbation method as been nonrenormalisable in previous studies, we cannot conclude that quantum general relativity itself is nonrenormalisable\footnote{See section 5 in Ref.\cite{nakanishi1990covariant}}.
Notably, coupling constant $c_{gr}$ was not considered in the previous studies for quantum general relativity, and a coupling constant absorbs infinitely large correction after a charge renormalization in the quantum Yang--Mills theory.
A role of the gravitational coupling constant in perturbative renormalization of general relativity is not fully investigated yet.
Llewellyn Smith suggested that renormalizability and tree-level unitarity of the theory are equivalent\cite{LLEWELLYNSMITH1973233}, as confirmed for some concrete examples; no exceptions has been reported to date.
A method to solve quantum general relativity in the Heisenberg picture was discussed by Nakanishi and Abe and summarized in Ref.\cite{doi:10.1143/PTP.111.301}.

Therefore, there are possibilities that a renormalisable perturbation of quantum general relativity exists along with their method.
Nonperturbative renormalizability of the proposed theory is not proven yet in this study; and further investigation is necessary such that, e.g., confirmation of an asymptotic safety using nonperturbative renormalization group methods\cite{Delamotte:2007pf, Dupuis:2020fhh}.

\subsection{Comparison with Chern--Simons gravity in the $(1\hspace{-.1em}+\hspace{-.1em}2)$-dimension}
It is known that Chern--Simons gravity in the $(1\hspace{-.1em}+\hspace{-.1em}2)$-dimension is a renormalisable theory.
The CWG theory in the $(1\hspace{-.1em}+\hspace{-.1em}3)$-dimension has a structure with that is similar to that of Chern--Simons gravity; thus, a comparison between these two theories may provide a deep insight into that renormalizability of the CWG theory.
The discussion of the Chern--Simons theory is provided by Witten in Ref.\cite{Witten198846}.

The Lagrangian of the $(1\hspace{-.1em}+\hspace{-.1em}2)$-dimensional Chern--Simons theory is provided as follows:
\begin{eqnarray}
\LLL_{C\hspace{-.1em}S}&:=&\frac{1}{2\hbar}\tilde\epsilon_{\bcdots\bcdot}\left(
d\tilde\www^\bcdots+\tilde\www^\bcdot_{~\star}\wedge\tilde\www^{\star\bcdot}\right)
\wedge\tilde\eee^\bcdot,\label{CSL}
\end{eqnarray}
where $\tilde\bullet$ shows a three-dimensional object corresponding to $\bullet$ in four dimensions.
Here the cosmological constant is omitted.
If this theory is renormalisable by power counting, $\tilde\eee$ and $\tilde\www$ must have an energy dimension; thus, the short-distance limit as a renormalization point for the perturbative expansion is provided as $\tilde\eee=\tilde\www=0$.
This trivial solution corresponds to the ``unbroken phase'' in the quantum field theory and is essential.
The perturbative expansion around the classical solution $\tilde\eee=\tilde\www=0$ is understandable in a three-dimensional case because the Lagrangian provided in (\ref{CSL}) has an invariant quadratic form under the $I\hspace{-.1em}SO(1,2)$ symmetry.
This opportunity in a three-dimensional space is accidental and independent from a lack of dynamical degree of freedom in the three-dimensional gravity.
On the other hand, the Einstein--Hilbert Lagrangian in four dimensions does not have either the Poincar\'{e} symmetry nor the quadratic invariant.
Witten has pointed out that:
``{\it As $e$ and $\omega$ have posivite dimension, the short-distance limit must have $e\hspace{-.1em}=\hspace{-.1em}\omega\hspace{-.2em}=\hspace{-.2em}0$. The problem is now that as eq.(3.13) has no quadratic term in an expansion around $e\hspace{-.1em}=\hspace{-.1em}\omega\hspace{-.2em}=\hspace{-.2em}0$, one cannot make sense of the ``unbroken phase'' that should govern the short-distance behavior; that is the essence of the unrenormalizability of quantum gravity in four dimensions.}'' (section 3.3 in Ref.\cite{Witten198846}).
Here, ``$e$'' and ``$\omega$'' correspond to $\eee$ and $\www$, and {\it eq.(3.13)} in Ref.\cite{Witten198846} is the four-dimensional Einstein--Hilbert Lagrangian, which is the same as equation (\ref{Lagrangian44}) with $\Lambda_c=0$ and $\cGR=1$ in this study.

Despite the absence of the Poincar\'{e} symmetry in the four-dimensional Einstein--Hilbert Lagrangian,
the Lagrangian has the co-Poincar\'{e} symmetry, and invariant quadratics exist as shown in section \ref{symplectic structure}.
The {\it essential} obstruction for the renormalizability of quantum gravity does not exist in the CWG theory.

\subsection{Comparison with the BF theory}\label{BFtheory}
The BF theory was first introduced by Pleba\'{n}ski in 1977\cite{doi:10.1063/1.523215}, although the term ``BF theory'' was not used at that time.
In 1989, Horowitz first treated the general relativity of the BF theory as a topological theory in the general $n$-dimensional spacetime manifold\cite{Horowitz1989}.
The BF theory has been reviewed in the literatures\cite{1991PhR...209..129B,Krasnov2011}.
The CWG theory investigated in this study is completely different from the BF theory presented in Ref.\cite{2018kurihara2}.
The essential difference between two theories are summarized below.

The BF theory is constructed on a four-dimensional spacetime Riemannian manifold $\M$ with $SO(1,3)$ (or $SO(4)$) symmetry.
The connection one-form and curvature two-form are obtained as the spin form $\www^{ab}$ and curvature two-form $\RRR^{ab}$, respectively, which are themselves the same as those of the CWG theory.
In addition, a new Lie-algebra value two-form $\BBB^{ab}$ is introduced; then the BF topological action is introduced as
\begin{eqnarray*}
\tilde{\I}_{BF}&=&\frac{1}{2}\int_{{\Sigma}_4}\epsilon_{\bcdots\bcdots}~\BBB^\bcdots\wedge\RRR^\bcdots,
\end{eqnarray*}
where ${\Sigma}_4$ is an appropriate four-dimensional manifold, which is simply connected and orientable.
The form $\BBB^{ab}$ is understood as a connection form on a $2$-bundle and forms the principal $2$-bundle in higher-gauge theory\cite{doi:10.1063/1.1790048}.
Action $\tilde{\I}_{BF}$ is topological and occurs according to the $2$-gauge theory;  thus, it is not equivalent to the Einstein--Hilbert action.
Some additional constraints\cite{doi:10.1063/1.523215} are necessary for applying the BF theory to general relativity.
This constraint in four-dimensional manifold, termed the simplicity condition, has been discussed by Gielen and Oritti\cite{Gielen:2010cu} as a linear constraint and by Celada, Gonz\'{a}lez, and Montesinos\cite{0264-9381-33-21-213001} as a constraint on $\CC$ formalism.

To convert the BF theory into gravitational theory, additional constraints must be implemented, for example,  a Lagrange multiplier term\cite{0264-9381-16-7-303} such as
\begin{eqnarray*}
{\I}_{BF}&=&\frac{1}{2}\int_{{\Sigma}_4}\left(
\epsilon_{\bcdots\bcdots}~\BBB^\bcdots\wedge\RRR^\bcdots
-\frac{1}{2}\phi_{\bcdots\bcdots}\BBB^\bcdots\wedge\BBB^\bcdots
\right),
\end{eqnarray*}
where $\phi_{abcd}$ is a scalar symmetric traceless matrix.
The simplicity condition appears as the Euler--Lagrange equation of motion with respect to $\phi$.
After implementing the constraint term, the action $\I_{BF}$ no longer has a characteristic class.
The simplicity condition $\epsilon_{ab\bcdots}\BBB^\bcdots=\SSS_{ab}$ is obtained as one of the solutions of the equation of motion.
This equivalence between the BF and CWG theories is valid as a solution of the equation of motion (on-shell condition) at a classical level, and the on-shell condition cannot be simply true after their quantization.
Therefore, the quantization of BF gravitational theory is complicated\cite{0264-9381-33-21-213001}.
Compared with the BF theory, the surface form is a topological object and one of the fundamental forms in the CWG theory.
Its definition (\ref{surface}) is valid at the operator level.

\section{Summary}\label{summary}
Herein, a canonical quantization based on canonical variables is performed in a Heisenberg picture using the Nakanishi--Kugo--Ojima formalism.
Because general relativity is a constrained system with global $GL(4,\R)$ and local $SO(1,3)$ symmetries, it is necessary to eliminate the unphysical d.o.f\@. by implementing gauge fixing and a ghost Lagrangian.
Even after gauge fixing, BRST symmetry remains.
We provided a complete set of quantum Lagrangian and BRST transformation operators with auxiliary and ghost fields in a self consistent manner.
Next, an appropriate Hilbert space and physical states are introduced into the theory.
Using the Kugo--Ojima theorem, this study shows the positivity of the physical states and the unitarity of the transition matrix.
The results obtained suggest that the renormalizability of quantum general relativity is worth reconsidering.

Although a quantization of pure gravitational theory without any matter nor gauge fields is discussed in this report, simultaneous quantization of general relativity and the Yang--Mills theory does not have any essential obstructions\cite{nakanishi1990covariant}.
For example, it is known that  a Dirac spinor has a gravitational  interaction through the covariant Dirac derivative as\cite{fre2012gravity} 
\begin{eqnarray*}
\gamma^\bcdot\partial_\bcdot\psi&\rightarrow&\gamma^\bcdot\left(\partial_\bcdot
+\frac{i}{4}c_{gr}\hspace{.1em}\Varepsilon_\bcdot^\mu\omega_\mu^{~\stars}\sigma_\stars
\right)\psi,
\end{eqnarray*}
where  $\sigma_{ab}:=i\eta_{a\bcdot}\eta_{b\bcdot}[\gamma^\bcdot,\gamma^\bcdot]/2$.
It is an interesting question if coupling constant $c_{gr}$ can absorb infinite quantum corrections due to gravitational interaction.

A term with the cosmological constant in the gravitational Lagrangian defined as
\begin{eqnarray*}
\LLL_\Lambda:=-\frac{\Lambda_c}{\hbar\kappa}\vvv=
\frac{\Lambda_c}{\hbar\kappa}\frac{1}{4!}\epsilon^{\bcdots\bcdots}\SSS_\bcdots\wedge\SSS_\bcdots,
\end{eqnarray*}
is not invariant under the co-Poincar\'{e} transformation; and thus, it is set to zero in this study.
In a context of quantum gravity, a quantum fluctuation of a vacuum energy may induced the cosmological constant as an effective field in the classical Lagrangian.
This possibility is discussed by the author in Ref.\cite{Kurihara_2018}.

\section*{Acknowledgment}
I appreciate the kind hospitality of all members of the theory group of Nikhef, particularly Prof. J. Vermaseren and Prof. E. Laenen.
A major part of this study was conducted during my stay at Nikhef in 2017.
I would also  like to thank Dr.~Y.~Sugiyama for his continuous encouragement and fruitful discussions.

The author would like to thank Enago for the English language review.
%
%
\vspace{1cm}
\appendix
\noindent
{\huge\bf Appendix:}\\
\vspace{-0.5cm}
\section{Co-translation operator and co-Poincar\'{e} symmetry}\label{cotrans}
In this Appendix, two remarks related to co-translation symmetry are given based on previous studies\cite{doi:10.1063/1.4990708, Kurihara_2020} by the author.
They are essentially important to construct the Chern--Weil gravitational theory introduced in this study.

Suppose $P_a$ be a generator of translation group $T^4$ on four-dimensional local Lorentz manifold $\M$.
Translation operator $\delta_T$ acts on vierbein and spin forms, respectively, as\cite{0264-9381-29-13-133001}
\begin{eqnarray*}
\delta_T\eee^a=d\xi^a+\www^a_{~\bcdot}\xi^\bcdot=d_\www\xi^a&{\rm and}&
\delta_T\www^{ab}=0,
\end{eqnarray*}
where $\xi^a$ is a local vector to characterize the translation.
Zanelli showed the Einstein--Hilbert Lagrangian in a three-dimensional spacetime had a translation invariant owing to an accidental opportunity in the three dimensional space\cite{0264-9381-29-13-133001}.
The co-translation is introduced to realize an extension of a translation symmetry in the Einstein--Hilbert Lagrangian in a four-dimensional spacetime. 
 
A generator of the co-translation is defined using the translation and contraction operators as follows:
\begin{definition}{\bf (co-translation)}: Co-translation generator $P_{ab}$ is defined as 
\begin{eqnarray*}
P_{ab}&=&P_a\iota_b,
\end{eqnarray*}
where $\iota_\bullet$ is a contraction operator with respect to local vector $\xi^\bullet$ in $\TM$, whose
representation is provided using the trivial basis as
\begin{eqnarray*}
\iota_{a}&=&\iota_{\xi^a},~~\xi^a=\eta^{ab}\Varepsilon^{\mu}_b\partial_\mu.
\end{eqnarray*}
The co-translation operator is represented as  $\delta_{CT}=\xi^a\times\delta_T\hspace{.1em}\iota_a$, where ``$\times$'' is an antisymmetric tensor product of the two vectors such that
\begin{eqnarray*}
\xi^a\times\xi^b=\xi^a\otimes\xi^b-\xi^b\otimes\xi^a=-\xi^b\times\xi^a
\end{eqnarray*}
\end{definition}
\noindent
A contraction is a map $\iota:\Omega^p(\TsM) \rightarrow \Omega^{p-1}(\TsM)$.
One can obtain a relation,
\begin{eqnarray*}
\iota_{a}\eee^b=\Varepsilon^{\mu}_a\Varepsilon_\nu^b\delta^\nu_\mu=\delta^b_a,&{\rm and}&
\delta_{CT}=\xi^\bcdot\hspace{-.1em}\times\delta_T\left(\iota_\bcdot\hspace{.1em}\eee^a\right)=
\xi^\bcdot\hspace{-.1em}\times\left(\delta_T\delta^a_\bcdot\right)=\xi^a,
\end{eqnarray*}
where $\delta_T\delta^a_b=\delta^a_b$ is used owing to its translation invariance in $\TM$.
Above mentioned relation is independent of the choice of bases in $\TM$.
A projection manifold and a projection bundle can be introduced  using a contraction operator as follows:
let $\M_{\perp a}\subset\M$ be a three-dimensional submanifold of $\M$, such that $\iota_a\aaa=0$ for $\aaa$ with $\Omega^1(\M_{\perp a})\ni\aaa\neq0$.
The trivial frame bundles in $T\M_{\perp a}$ and $T^*\M_{\perp a}$ are regarded as sub-bundles of $T\M$ and $T^*\M$, respectively.
If $\M$ has Poincar\'{e} symmetry $I\hspace{-.1em}S\hspace{-.1em}O(1,3)$, the submanifold $\M_{\perp a}$ has the $I\hspace{-.1em}S\hspace{-.1em}O(1,2)$ or the $I\hspace{-.1em}S\hspace{-.1em}O(3)$ symmetry.
Therefore, connection $\AAA_a$ and curvature $\FFF_a=d\AAA_a+\AAA_a\wedge\AAA_a$ are provided in $\M_{\perp a}$ as sub-bundles of $\AAA$ and $\FFF$ on $\M$, respectively.
Using an equivalence relation $\sim_a$ such as $\aaa\sim_a\bbb \Leftrightarrow \iota_a\aaa=\iota_a\bbb$, quotient bundles $\widetilde{\AAA}_a=\AAA/{\sim_a}$ and $\widetilde{\FFF}_a=\FFF/{\sim_a}$ can be introduced, where $\aaa$ and $\bbb$ are any $p$-forms on $T^*\M$.
It is clear that $\AAA_a\simeq\widetilde{\AAA}_a\subset\AAA$ and $\FFF_a\simeq\widetilde{\FFF}_a\subset\FFF$.

The co-translation acts on the fundamental forms as follows:
\begin{eqnarray}
\delta_{CT}\left(\SSS_{ab}\right)&=&\frac{1}{2}\epsilon_{abcd}\left(
\xi^c\times\td\xi^d-\xi^d\times\td\xi^c,\label{dctS}\right)\\
\delta_{CT} \www^{ab}&=&0.\label{dctW}
\end{eqnarray}

\begin{remark}\label{remA2}
{\rm
({\bf Remark 3.2} in Ref.\cite{doi:10.1063/1.4990708})}\\
The Einstein--Hilbert Lagrangian given in (\ref{Lagrangian44}) with $\Lambda_c=0$ is invariant under the co-Poincar\'{e} transformation up to a total derivative.
\end{remark}
\begin{proof}
The gravitational Lagrangian with $\Lambda_c=0$ is considered.
The invariance under the local $SO(1,3)$ symmetry of the Einstein--Hilbert Lagrangian is trivial from its definition.
A co-translation operator $\delta_{CT}$ induces a transformation on the Einstein--Hilbert Lagrangian $\LLL_G$ as
\begin{eqnarray}
(\hbar\kappa)\delta_{CT}\LLL_G=\delta_{CT}\left(\frac{1}{2}
\epsilon_{\bcdots\bcdots}\RRR^\bcdots\wedge\eee^\bcdot\wedge\eee^\bcdot
\right)=
\epsilon_{\bcdots\bcdots}\RRR^\bcdots\wedge\left(\xi^\bcdot\times\td\xi^\bcdot\right)
=\td\left\{
\epsilon_{\bcdots\bcdots}\RRR^\bcdots(\xi^\bcdot\times\xi^\bcdot)
\right\},\label{deltactL}
\end{eqnarray}
where (\ref{dctS}), (\ref{dctW}) and the Bianchi identity $\td\RRR=0$ are used.
Owing to the definition of the covariant derivative (\ref{connec2}), $\td\{\cdots\}=d\{\cdots\}$ is maintained in (\ref{deltactL}); thus the co-translation operator only induces a total derivative term on the Einstein--Hilbert Lagrangian without the cosmological constant.
\end{proof}

\begin{remark}\label{remA3}
{\rm
({\bf Theorem 4.2} in Ref.\cite{doi:10.1063/1.4990708} and {\bf Remark 2.1} in Ref.\cite{Kurihara_2020})}\\
The Einstein--Hilbert Lagrangian without the cosmological constant is defined using the curvature of the co-Poincar\'{e} bundle as (\ref{cPL}).
It has a topological invariant as the second Chern class $c_2(\FFF_\cP):$ 
\begin{eqnarray*}
\LLL=
\frac{8\pi^2}{\kappa}c_2(\FFF_\cP)\in \R\otimes H^4(\M,\Z).
\end{eqnarray*}
\end{remark}
\begin{proof}
The right-hand side of (\ref{cPL}) is expressed using the structure constants as
\begin{eqnarray*}
\Tr\left[\FFF_\cP\wedge\FFF_\cP\right]&:=&\frac{1}{2}
\FF^K_{~IJ}\hspace{.1em}\FFF_\cP^I\wedge\FFF_\cP^J\hspace{.1em}T_K. 
\end{eqnarray*}
Owing to the definition of the co-Poincar\'{e} curvature (\ref{cur2}) and the Lie algebra of the co-Poincar\'{e} symmetry, a direct calculation with a trivial basis yields
\begin{eqnarray*}
\Tr\left[\FFF_\cP\wedge\FFF_\cP\right]&=&
\Tr\left[\RRR^{ac}\wedge
d_\www\left(P_{a}^{\hspace{.3em}b}\hspace{.1em}\SSS_{cb}\right)\right]
=
d_\www\hspace{-.2em}\left(\Tr\left[P_{a}^{\hspace{.3em}b}\right]
\RRR^{ac}\wedge\SSS_{cb}\right)=
-d_\www\hspace{-.2em}
\left(P_{\bcdot}^{\hspace{.3em}\bcdot}
\hspace{.1em}\RRR\wedge\SSS\right),
\end{eqnarray*}
where Bianchi identity $d_\www\RRR=0$ and co-translation invariance $P\hspace{.1em}(\RRR)=0$ are used.
For the last equality, we note that $\Tr\left[\RRR^{ac}\wedge\SSS_{cb}\right]=-\delta^a_b(\RRR\wedge\SSS)$.
The Einstein--Hilbert gravitational Lagrangian is co-translation-invariant up to total derivative as given as {\bf Remark \ref{remA2}}; thus, $P$ is removed from the last expression and it is embedded in a five-dimensional line bundle of $\MM_5:=\MM\otimes\R$, whose boundary is given as $\partial\MM_5=\Sigma$.
Consequently,  
\begin{eqnarray*}
-\int_{\Sigma}\Tr\left[\FFF_\cP\wedge\FFF_\cP\right]&=&
\int_{\MM_5}d_\www\left(\RRR\wedge\SSS\right)=
\int_{\MM_5}d\left(\RRR\wedge\SSS\right)=
\int_{\partial\hspace{-.1em}\MM_5=\Sigma}\RRR\wedge\SSS.
\end{eqnarray*}
A surface term is eliminated owing to a boundary condition.
We note that 
\begin{eqnarray*}
d_\www\left(\RRR\wedge\SSS\right)=
d\left(\RRR\wedge\SSS\right)
+\www\wedge\RRR\wedge\SSS
-\RRR\wedge\SSS\wedge\www=d\left(\RRR\wedge\SSS\right),
\end{eqnarray*}
owing to the definition of a covariant derivative.

The second Chern class with respect to the co-Poincar\'{e} curvature is obtained as  
\begin{eqnarray*}
c_2(\FFF_\cP)&:=&\frac{1}{8\pi^2}\left(\Tr\left[\FFF_\cP\right]^2
-\Tr\left[\FFF_\cP\wedge\FFF_\cP\right]\right),
\end{eqnarray*}
and it has a homology class as $c_2(\FFF_\cP)\in H^4(\M,\R)\subset\mathrm{Im}\hspace{.1em}H^4(\M,\Z)$ owing to the Chern--Weil theory.
The co-Poincar\'{e} curvature $\FFF_\cP$ is traceless; thus the remark is maintained.
\end{proof}
\noindent
This result suggests that appropriate fundamental forms of the symplectic geometry for general relativity can be identified as $(\www,\SSS)$.
This choice of the canonical pair of general relativity is equivalent to that proposed by Kanatchikov\cite{Kanatchikov_2013} on the basis of the de\hspace{.1em}Donder--Weyl Hamiltonian theory.

\section{Proof of nilpotency}\label{ap1}
\noindent\fbox{{\bf Coordinate vector}}\\
The coordinate vectors are fundamental vectors on $T\MM$. 
Nilpotent can be confirmed as
\begin{eqnarray*}
&~&
\delBRST\left[\delBRST\left[x^{\mu}\right]\right]
~=~\delBRST\left[\chi^\mu\right]
~=~\delBRST\left[g^{\mu\nu}\chi_\nu\right],\\
&~&~=~
g^{\mu\rho}\left(\partial_\rho\chi^\nu\right)\chi_\nu+
g^{\rho\nu}\left(\partial_\rho\chi^\mu\right)\chi_\nu-g^{\mu\nu}g_{\nu\rho}\left(
\partial^\rho\chi^\sigma+
\partial^\sigma\chi^\rho
\right)\chi_\sigma,\\
&~&~=~
\left(\partial^\mu\chi^\nu\right)\chi_\nu+
\left(\partial^\nu\chi^\mu\right)\chi_\nu-\left(\partial^\mu\chi^\sigma\right)\chi_\sigma-
\left(\partial^\sigma\chi^\mu\right)\chi_\sigma=0.
\end{eqnarray*}
\noindent
\fbox{{\bf Metric tensor}}\\
Starting from the BRST transformation of the metric tensor (\ref{BRSTg}), nilpotent is provided as
\begin{eqnarray*}
\delBRST\left[\delBRST\left[g_{\mu\nu}\right]\right]&=&
\delBRST\left[
-g_{\mu\rho}\partial_\nu\chi^\rho
\right]
+\delBRST\left[\mu\leftrightarrow\nu\right],\nonumber\\
&=&\left\{-\delBRST\left[g_{\mu\rho}\right]\partial_\nu\chi^\rho
+g_{\mu\rho}\left(\partial_\nu\delBRST\left[x^\sigma\right]\right)\partial_\sigma\chi^\rho\right\}
+\left\{\mu\leftrightarrow\nu\right\},\nonumber\\
&=&\left\{
g_{\mu\sigma}\left(\partial_\rho\chi^\sigma\right)\partial_\nu\chi^\rho+
g_{\mu\rho}\left(\partial_\nu\chi^\sigma\right)\partial_\sigma\chi^\rho\right\}
+\left\{\mu\leftrightarrow\nu\right\}=0,
\end{eqnarray*}
where anticommutativity of the ghost filed is used.\\

\noindent
\fbox{{\bf Ghost field}}\\
Since the ghost field has two parts, nilpotent is checked separately.
First, nilpotent of the $\chi_\mu$ is shown as 
\begin{eqnarray*}
\delBRST\left[\delBRST\left[\chi_\mu\right]\right]
&=&
\delBRST\left[\delBRST\left[g_{\mu\nu}\chi^\nu\right]\right]
~=~\delBRST\left[\delBRST\left[g_{\mu\nu}\right]\right]\chi^\nu=0,
\end{eqnarray*}
where $\delBRST[\chi^\mu]=0$ and nilpotent of the metric tensor are used.
Direct calculation from (\ref{BRSTchi}) gives the same result, too.
The second part becomes
\begin{eqnarray*}
\delBRST\left[\delBRST\left[\chi^a_{~b}\right]\right]&=&
\delBRST\left[\chi^a_{~c}\chi^{c}_{~b}\right]
~=~
\chi^a_{~c_2}\chi^{c_2}_{~~c_1}\chi^{c_1}_{~~b}-\chi^a_{~c_1}\chi^{c_1}_{~~c_2}\chi^{c_2}_{~~b}=0,
\end{eqnarray*}
due to anticommutativity of the ghost field.

Next, a tensor $\partial_{\mu}\chi^{\nu}$ is also nilpotent as
\begin{eqnarray*}
\delBRST\left[\delBRST\left[\partial_{\mu}\chi^{\nu}\right]\right]&=&
-\delBRST\left[\partial_{\mu}\chi^{\rho}\partial_{\rho}\chi^{\nu}\right]=~
-\partial_{\mu}\chi^{\rho_1}\partial_{\rho_1}\chi^{\rho_2}\partial_{\rho_2}\chi^{\nu}
+\partial_{\mu}\chi^{\rho_1}\partial_{\rho_1}\chi^{\rho_2}\partial_{\rho_2}\chi^{\nu}=~0.
\end{eqnarray*}
\noindent
\fbox{{\bf Vierbein form}}\\
Nilpotent of the vierbein form is provided as
\begin{eqnarray*}
\delBRST\left[\delBRST\left[\eee^a\right]\right]&=&
\delBRST\left[\eee^b\chi^a_{~b}\right]=~
\eee^{b_1}\chi^{b_2}_{~~b_1}\chi^a_{~b_2}+\eee^{b_2}\chi^a_{~b_1}\chi^{b_1}_{~~b_2}=0,
\end{eqnarray*}
due to anticommutativity of the ghost field.\\

\noindent
\fbox{{\bf Spin form}}\\
One can trace the same calculation as a case of the vierbein form due to $\delBRST\left[d\chi^{ab}\right]=0$.
Detailed calculations are omitted here.\\

\noindent
\fbox{{\bf Volume form}}\\
The volume form is global scalar and their BRST transformation is expected to vanish, which can be confirmed as
\begin{eqnarray*}
\delBRST\left[\vvv\right]&=&\frac{1}{4!}\epsilon_{\bcdott}
\delBRST\left[\eee^\bcdot\wedge\eee^\bcdot\wedge\eee^\bcdot\wedge\eee^\bcdot\right]~=~
\frac{1}{3!}\epsilon_{a_1\bcdot\bcdots}
\delBRST\left[\chi^{a_1}_{~~a_2}\eee^{a_2}\wedge\eee^\bcdot\wedge\eee^\bcdot\wedge\eee^\bcdot\right]~=~0,
\end{eqnarray*}
due to $\eee^\bcdot\wedge\eee^\bcdot\wedge\eee^\bcdot\wedge\eee^\bcdot\propto\epsilon^{\bcdott}$ and $\chi^{a_1}_{~~a_2}=0$ when $a_1=a_2.$\\

\noindent
\fbox{{\bf Surface form}}\\
The BRST transformation of the surface form is provided as
\begin{eqnarray*}
\delBRST\left[\SSS_{ab}\right]&=&\frac{1}{2}\epsilon_{abc_1c_2}\delBRST\left[\eee^{c_1}\wedge\eee^{c_2}\right]
=~\epsilon_{abc_1c_2}\chi^{c_1}_{~c_3}\eee^{c_3}\wedge\eee^{c_2}~
\left(=~\ccc^\bcdot\wedge\overline{\eee}_{ab\bcdot}\right).
\end{eqnarray*}
Applying the BRST-transformation on it again, one can get
\begin{eqnarray*}
&~&\delBRST\left[\delBRST\left[\SSS_{ab}\right]\right]~=~
\epsilon_{abc_1c_2}\delBRST\left[\chi^{c_1}_{~c_3}\eee^{c_3}\wedge\eee^{c_2}\right],\\
&~&~=~\epsilon_{abc_1c_2}\Bigl\{
\chi^{c_1}_{~c_4}\chi^{c_4}_{~c_3}\eee^{c_3}\wedge\eee^{c_2}-
\chi^{c_1}_{~c_3}\chi^{c_3}_{~c_4}\eee^{c_4}\wedge\eee^{c_2}-
\chi^{c_1}_{~c_3}\chi^{c_2}_{~c_4}\eee^{c_3}\wedge\eee^{c_4}
\Bigr\}=0,
\end{eqnarray*}
because first term is the same as the second term and the third term is symmetric with $c_1$ and $c_2$ exchange.\\

\noindent
\fbox{{\bf ghost forms}}\\
The BRST transformation of $\ccc^a$ is given by
\begin{eqnarray*}
\delBRST\left[\ccc^a\right]&=&
\delBRST\left[\chi^a_{~b}~\Varepsilon^b_\mu~dx^\mu\right],\\
&=&\chi^a_{~b_1}\chi^{b_1}_{~b_2}\Varepsilon^{b_2}_\mu dx^\mu-
\chi^a_{~b_1}~\Varepsilon^{b_2}_\mu \chi^{b_2}_{~b_1}dx^\mu
+\chi^a_{~b}~\left(\partial_\mu\chi^\nu\right)\Varepsilon^b_\nu~dx^\mu
-\chi^a_{~b}~\Varepsilon^b_\mu d\chi^\mu=0,
\end{eqnarray*} 

\noindent
\fbox{{\bf Other forms}}\\
Nilpotent of other forms are trivial and the proof is omitted here.\\

\vskip 2mm
\noindent
\fbox{{\bf Gravitational Lagrangian}}\\
The quantum Lagrangian must be the BRST-null.
Gauge-fixing and Fadeef--Popov Lagrangians are constructed to satisfy the BRST-null condition in section {6-2}.
Nilpotent of only the gravitational Lagrangian is given here.
The BRST transformation of the gravitational Lagrangian is provided as
\begin{eqnarray*}
\delBRST\left[\LLL_G\right]&=&
\frac{1}{2}\delBRST\left[\left(
d\www^\bcdots
+\cGR\hspace{.1em}\www^\bcdot_{~\star}\wedge\www^{\star\bcdot}\right)\wedge{\SSS}_\bcdots
-\frac{\Lambda}{3!}\vvv\right].
\end{eqnarray*}
The BRST transformation for the volume form is vanished by itself. 
For the derivative term,
\begin{eqnarray*}
\delBRST\left[
d\www^\bcdots\wedge{\SSS}_\bcdots
\right]&=&
\epsilon_{abc_2c_3}\chi_{~c_1}^{b}d\www^{ac_1}\wedge\eee^{c_2}\wedge\eee^{c_3}+
\epsilon_{abc_2c_3}\www^{ac_1}\wedge d\chi_{~c_1}^{b}\wedge\eee^{c_2}\wedge\eee^{c_3}\\&~&+
\epsilon_{abc_1c_2}\chi^{c_1}_{~c_3}~d\www^{ab}\wedge\eee^{c_3}\wedge\eee^{c_2}\\
&=&2~\www^{ac_1}\wedge d\chi^{b}_{~c_1}\wedge\SSS_{ab},
\end{eqnarray*}
where first- and third-terms are cancelled each other.
Remnant term is transformed as
\begin{eqnarray*}
\delBRST\left[\www^\bcdot_{~\star}\wedge\www^{\star\bcdot}\wedge{\SSS}_\bcdots\right]&=&
\epsilon_{abc_2c_3}\chi^{c_2}_{~c_4}\www^{ac_1}\wedge\www_{c_1}^{~~b}\wedge\eee^{c_4}\wedge\eee^{c_3}+
\epsilon_{abc_3c_4}\chi^{c_2}_{~c_1}\www^{ac_1}\wedge\www_{c_2}^{~~b}\wedge\eee^{c_3}\wedge\eee^{c_4}\\&~&+
\epsilon_{abc_3c_4}\chi^{b}_{~c_2}\www^{ac_1}\wedge\www_{c_1}^{~~c2}\wedge\eee^{c_3}\wedge\eee^{c_4}-2{\cGR}^{-1}\www^{ac_1}\wedge d\chi^b_{~c_1}\wedge\SSS_{ab},\\
&=&-2{\cGR}^{-1}\www^{ac_1}\wedge d\chi^b_{~c_1}\wedge\SSS_{ab}.
\end{eqnarray*}
In the r.h.s of the first line, the second term is zero as itself, and first- and third-terms are cancelled each other.
Therefore one can confirmed $\delBRST\left[\LLL_G\right]=0$ after summing up all terms.

If we use a following remake, we can give simpler proofs for above forms.\\
\noindent
\fbox{{\bf Remark}}\\
If both of two fields, $\alpha$ and $\beta$, are nilpotent, $\alpha\beta$ is also nilpotent.\\
\noindent
{\it Proof:}\\
If a field $X$ is nilpotent, signatures of the Leibniz rule satisfy $\epsilon_{X}=-\epsilon_{\delta X}$ due to $\delBRST[\delBRST[X]]=0$ and (\ref{Leib}), where
$\epsilon_{X}$ ($\epsilon_{\delta X}$) is a signature of $X$ ($\delBRST[X]$), respectively.
Therefore
\begin{eqnarray*}
\delBRST\left[\delBRST\left[\alpha\beta\right]\right]&=&
\epsilon_{\alpha}\delBRST\left[\alpha\right]\delBRST\left[\beta\right]+
\epsilon_{\delta\alpha}\delBRST\left[a\right]\delBRST\left[\beta\right]
~=~0.
\end{eqnarray*} 

\section{Equations of motion}\label{app2}
From the classical Lagrangian form, the torsionless condition and the Einstein equation are obtained as equations of motion by requiring a stationary condition for variation of action.
The same procedure can be used to obtain Euler--Lagrange equations from the quantum Lagrangian.
Here, the quantum Lagrangian is summarized as
\begin{eqnarray*}
&~&\LLLQG=\LLL_G+\LLLGF+\LLLFP,~~~~~~~~~~~~~~~~~~~~~~~~~~~~~~~~~~~~~~~(\ref{QLGF})\\
&~&
\left\{
\begin{array}{llc}
\LLL_G&=&~~~~~~\frac{1}{2}\left(
\RRR^\bcdots
-\frac{\Lambda}{3!}\overline{\SSS}^\bcdots\right)\wedge{\SSS}_\bcdots,~~~~~~~~~~~~~~~~~~~~~~(\ref{Lagrangian44})\\
\LLLGF&=&
-\frac{1}{2}\left(d\bbb^\bcdots+\alpha\bbb^\bcdot_{~\star}\wedge\bbb^{\star\bcdot}
\right)\wedge\SSS_\bcdots,~~~~~~~~~~~~~~~~~~~~(\ref{LGF})\\
\LLLFP&=&
-\frac{i}{2}\left(
d\tilde{\ccc}^\bcdots+
\alpha
\tilde{\ccc}^{\bcdot}_{~\star}\wedge\bbb^{\star\bcdot}\right)
\wedge\ccc^\star\wedge\overline{\eee}_{\star\bcdots},~~~~~~~~~~~~~~~(\ref{LFP})
\end{array}
\right.
\end{eqnarray*}
Euler--Lagrange equations are obtained as follows:

\noindent
\fbox{$\delta_{\www}$}
\begin{eqnarray*}
\TTT^a&=&d\eee^a+\cGR\hspace{.1em}\www^a_{~\bcdot}\wedge\eee^\bcdot=0.
\end{eqnarray*}
This is the torsionless condition in the the same form as the classical Lagrangian.\\
\noindent
\fbox{$\delta_{\eee}$}
\begin{eqnarray}
\frac{1}{2}\overline{\left(\RRR^\bcdots\wedge\eee^\bcdot\right)}_a-\Lambda\VVV_a
-\frac{1}{2}\left(d\bbb^\bcdots+\alpha\bbb^\bcdot_{~\star}\wedge\bbb^{\star\bcdot}\right)\wedge\overline{\eee}_{a\bcdots}
-\frac{i}{2}\left(d\tilde{\ccc}^\bcdots+\alpha\tilde{\ccc}^\bcdot_{~\star}\wedge\bbb^{\star\bcdot}\right)\wedge\overline{\ccc}_{a\bcdots}
&=&0,\label{ELE2}
\end{eqnarray}
where $\VVV_a=\epsilon_{ab_1b_2b_3}\eee^{b_1}\wedge\eee^{b_2}\wedge\eee^{b_3}/3!$.
The first two terms give the Einstein equation without matter fields. 
Third- and fourth-terms newly appeared from the gauge-fixing and Feddeev--Popov Lagrangian forms.\\
\noindent
\fbox{$\delta_{\bbb}$}
\begin{eqnarray*}
d\SSS_{ab}-\alpha\left(
\bbb^\bcdot_a\wedge\SSS_{\bcdot b}-i\tilde{\ccc}^\bcdot_a\wedge\ccc^\bcdot\wedge\overline{\eee}_{\bcdots b}
\right)&=&0.
\end{eqnarray*}
We note that $\bbb$ and $\delta_\bbb$ are anticommute each other and the variation operator is applied from the left.
When the Landau gauge is used, the de Donder gauge-fixing condition $d\SSS_{ab}=0$ is obtained.\\

\noindent
\fbox{$\delta_{\ccc}$}
\begin{eqnarray}
\left(d\tilde{\ccc}^\bcdots+
\alpha
\tilde{\ccc}^{\bcdot}_{~\star}\wedge\bbb^{\star\bcdot}\right)\wedge\overline{\eee}_{a\bcdot\bcdot}&=&0,\label{ELE3}
\end{eqnarray}
where the anticommutation among $\ccc$, $\delta\ccc$ and $\bbb$ is used.\\

\noindent
\fbox{$\delta_{\tilde{\ccc}}$}
\begin{eqnarray*}
\epsilon_{ab\bcdots}\left(
d\left(\ccc^\bcdot\wedge\eee^\bcdot\right)-\alpha
\bbb^\bcdot_{~\star}\wedge\ccc^\star\wedge\eee^\bcdot
\right)\nonumber&=&
\epsilon_{ab\bcdots}\left(
d\ccc^\bcdot-\alpha
\bbb^\bcdot_{~\star}\wedge\ccc^\star
\right)\wedge\eee^\bcdot
~=~0,
\end{eqnarray*}
where the de\hspace{.2em}Donder condition is used. 

The BRST transformation may give another set of equations, which must be consistent with above equations:
\begin{eqnarray*}
\delBRST\left[\TTT^a\right]&=&\chi^a_{~\bcdot}~d\eee^\bcdot+d\chi^a_{~\bcdot}\wedge\eee^\bcdot
+\cGR\hspace{.1em}\chi^a_{~\bcdot}~\www^{~\bcdot}_{\star}\wedge\eee^\star
+\cGR\hspace{.1em}\chi_{\star}^{~\bcdot}~\www^{a}_{~\bcdot}\wedge\eee^\star
-d\chi^a_{~\bcdot}\wedge\eee^\bcdot
+\cGR\hspace{.1em}\chi^\bcdot_{~\star}\www^a_{~\bcdot}\wedge\eee^\star
\nonumber\\
&=&\chi^a_{~\bcdot}~\TTT^\bcdot~=~0.
\end{eqnarray*}
This is consistent with the torsion-less condition.
The BRST transformation for the volume form is vanished and last two terms are cancelled each other such as
\begin{eqnarray*}
\delBRST\left[i\left(d\tilde{\ccc}^\bcdots+\alpha\tilde{\ccc}^\bcdot_{~\star}\wedge\bbb^{\star\bcdot}\right)\wedge\overline{\ccc}_{a\bcdots}\right]\nonumber&=&
-\left(d\bbb^\bcdots+\alpha\bbb^\bcdot_{~\star}\wedge\bbb^{\star\bcdot}\right)\wedge\overline{\eee}_{a\bcdots}
\end{eqnarray*}
Therefor, the BRST transformation of (\ref{ELE2}) is given by
\begin{eqnarray*}
0&=&
\epsilon_{abc\bullet}\delBRST\left[\left(d\www^{ab}
+\cGR\hspace{.1em}\www^a_{~\bcdot}\wedge\www^{\bcdot b}\right)\wedge\eee^c\right]\\
&=&\epsilon_{abc\bullet}\Bigl\{
\chi^b_{~\bcdot}\left(d\www^{a\bcdot}
+\cGR\hspace{.1em}\www^a_{~\star}\wedge\www^{\star\bcdot}\right)\wedge\eee^c
+
\left(d\www^{ab}+\cGR\hspace{.1em}\www^a_{~\bcdot}\wedge\www^{\bcdot b}\right)\wedge\ccc^c
\Bigr\}.
\end{eqnarray*}
This is consistent with the Einstein equation.
\begin{eqnarray*}
\delBRST\left[d\eee^a-\alpha\left(
\bbb^a_{~\bcdot}\wedge\eee^\bcdot-i\tilde{\ccc}_\bcdot^a\wedge\ccc^\bcdot
\right)\right]&=&d\ccc^a=0,
\end{eqnarray*}
where $\alpha$-terms are cancelled each other.
The BRST transformation of (\ref{ELE3}) gives an equation of motion for $\bbb$ in (\ref{ELE2}). 
\begin{eqnarray*}
\epsilon_{ab\bcdots}\delBRST\left[
d\left(\ccc^\bcdot\wedge\eee^\bcdot\right)
-\alpha
\bbb^\bcdot_{~\star}\wedge\ccc^\star\wedge\eee^\bcdot
\right]&=&0,
\end{eqnarray*}
where $\epsilon_{ab\bcdots}\ccc^\bcdot\wedge\ccc^\bcdot=0$ is used.
This is not an equation, but an identity.

\section{proof of (\ref{BRSTgen2})}\label{appC}
The proof of (\ref{BRSTgen2}) can be given as follows: 
Shorthand notations are introduced to omit indices in this appendix for simplicity as follows:
\begin{eqnarray*}
\widehat{\QQQ}_\bbb=\frac{1}{2}\widehat{\bm\beta}\bcdot\delBRST\left[\widehat{\bm\QQQ}\right],&~~&
\widehat{\QQQ}_{\tilde\ccc}=\frac{i}{4}\widehat{\bm\chi}\bcdot\delBRST\left[\widehat{\bm\QQQ}\right]
-\frac{1}{4}\widehat{\bm\beta}\bcdot\widehat{\bm\QQQ},\\
\end{eqnarray*}
where $\widehat{\bm\QQQ}$ is defined in equation (\ref{defQQQ}).  
By using these notions, the commutation relation can be represented as
\begin{eqnarray}
-8i\left[\widehat{\QQQ}_{\tilde\ccc},\widehat{\QQQ}_\bbb\right]&=&
\left[
\widehat{\bm\beta}\bcdot\delBRST\left[\widehat{\bm\QQQ}\right],
\widehat{\bm\chi}\bcdot\delBRST\left[\widehat{\bm\QQQ}\right]
\right]+i
\left[
\widehat{\bm\beta}\bcdot\delBRST\left[\widehat{\bm\QQQ}\right],
\widehat{\bm\beta}\bcdot \widehat{\bm\QQQ}
\right],\nonumber\\&=&
\widehat{\bm\beta}\left[\delBRST\left[\widehat{\bm\QQQ}\right], \widehat{\bm\chi}\right]\delBRST\left[\widehat{\bm\QQQ}\right]
+\widehat{\bm\chi}
\left[\widehat{\bm\beta},\delBRST\left[\widehat{\bm\QQQ}\right]
\right]\delBRST\left[\widehat{\bm\QQQ}\right]
\nonumber\\&~&
+i\hspace{.1em}
\widehat{\bm\beta}\left[\delBRST\left[\widehat{\bm\QQQ}\right],\widehat{\bm\beta}\bcdot\widehat{\bm\QQQ}\right]+i
\left[
\widehat{\bm\beta},
\widehat{\bm\beta}\bcdot\widehat{\bm\QQQ}
\right]\delBRST\left[\widehat{\bm\QQQ}\right].\label{DD}
\end{eqnarray}
where $[\widehat{\bm\beta},\widehat{\bm\chi}]=0$ is used.
The first term of (\ref{DD}) becomes
\begin{eqnarray*}
\widehat{\bm\beta}\left[\delBRST\left[\widehat{\bm\QQQ}\right], \widehat{\bm\chi}\right]\delBRST\left[\widehat{\bm\QQQ}\right]&=&\widehat{\bm\beta}
\left(
\delBRST\left[\widehat{\bm\QQQ}\right]\widehat{\bm\chi}-
\widehat{\bm\chi}~\delBRST\left[\widehat{\bm\QQQ}\right]
\right)\delBRST\left[\widehat{\bm\QQQ}\right],
\nonumber\\&=&
\widehat{\bm\beta}
\left(
\delBRST\left[
\left\{\widehat{\bm\QQQ},\widehat{\bm\chi}\right\}
\right]+i\left[\widehat{\bm\beta},\widehat{\bm\QQQ}\right]
\right)\delBRST\left[\widehat{\bm\QQQ}\right],\\&=&
i\widehat{\bm\beta}~
\left[\widehat{\bm\beta},\widehat{\bm\QQQ}\right]
\delBRST\left[\widehat{\bm\QQQ}\right]=~4\widehat{\QQQ}_\bbb,
\end{eqnarray*}
where $\delBRST\left[\left\{\widehat{\bm\QQQ},\widehat{\bm\chi}\right\}\right]=0$ due to (\ref{fCR3}), and (\ref{fCR2}) are used.
Note that $\widehat{\bm\chi}$ and $\widehat{\bm\QQQ}$ have $\epsilon_X=-1$ in (\ref{Leib}).
The second term of (\ref{DD}) is zero, because
\begin{eqnarray*}
\left[\widehat{\bm\beta},\delBRST\left[\widehat{\bm\QQQ}\right]
\right]&=&
\delBRST\left[
\left[
\widehat{\bm\beta},\widehat{\bm\QQQ}
\right]
\right]=~0,
\end{eqnarray*}
due to (\ref{fCR2}).
The third term is also zero as
\begin{eqnarray*}
i\widehat{\bm\beta}\left[\delBRST\left[\widehat{\bm\QQQ}\right],\widehat{\bm\beta}\bcdot\widehat{\bm\QQQ}\right]&=&
i\widehat{\bm\beta}\widehat{\bm\beta}
\left[\delBRST\left[\widehat{\bm\QQQ}\right],\widehat{\bm\QQQ}\right]+
i\widehat{\bm\beta}~
\delBRST\left[\left[\widehat{\bm\QQQ},\widehat{\bm\beta}\right]\right]\widehat{\bm\QQQ}=~0,
\end{eqnarray*}
due to (\ref{fCR2}).
A relation $\left[\delBRST\left[\widehat{\bm\QQQ}\right],\widehat{\bm\QQQ}\right]=0$ can be confirmed by direct calculations.
The last term of (\ref{DD}) becomes
\begin{eqnarray*}
i\left[
\widehat{\bm\beta},
\widehat{\bm\beta}\bcdot\widehat{\bm\QQQ}
\right]\delBRST\left[\widehat{\bm\QQQ}\right]&=&
i\widehat{\bm\beta}~
\left[\widehat{\bm\beta},\widehat{\bm\QQQ}\right]
\delBRST\left[\widehat{\bm\QQQ}\right]=~4\widehat{\QQQ}_\bbb.
\end{eqnarray*}
\bibliographystyle{elsarticle-num}
\bibliography{ref}
\end{document}